\documentclass[11pt,letterpaper]{article} 
\usepackage[letterpaper,margin=1in]{geometry}

\usepackage[UKenglish]{babel}
\usepackage{fullpage}
\usepackage{graphicx}
\usepackage{amsmath,amssymb,amsthm}
\usepackage{xcolor}
\usepackage{algorithm,algpseudocode}
\usepackage[colorlinks=true, allcolors=blue]{hyperref}
\usepackage[capitalise,noabbrev]{cleveref}
\usepackage{thmtools} 
\usepackage{booktabs,tabularx}
\usepackage{thm-restate}
\usepackage{enumerate}
\usepackage{tikz}
\usepackage{placeins}

\crefname{algorithm}{Algorithm}{Algorithms}

\newtheorem{theorem}{Theorem}
\newtheorem*{theorem*}{Theorem}
\newtheorem{lemma}{Lemma}
\makeatletter
\let\c@lemma\c@theorem
\makeatother

\makeatletter
\let\c@corollary\c@theorem
\makeatother

\title{Polylogarithmic Approximation for Packing Convex Polyhedra}

\author{Anders Aamand\thanks{\texttt{aa561@rice.edu}.}\\ Rice University \and Mikkel Abrahamsen\thanks{\texttt{miab@di.ku.dk}. Partially supported by the Independent Research Fund Denmark, grant 1054-00032B, and by the Carlsberg Foundation, grant CF24-1929.}\\ University of Copenhagen \and Simon Bartlmae\thanks{\texttt{simon.bartlmae7@gmail.com}. Partially supported by the Independent Research Fund Denmark, grant 1054-00032B, and by the Carlsberg Foundation, grant CF24-1929.}\\ University of Copenhagen}
\date{}

\begin{document}

\maketitle

\begin{abstract}

We study translational packing of convex polyhedra in three dimensions. Given $n$ convex polyhedra in $\mathbb{R}^3$, we seek a placement with pairwise disjoint interiors that minimizes the volume of an axis-parallel bounding box. Constant-factor approximation algorithms are known when rotations are allowed, but no approximation guarantee was previously known for arbitrary convex polyhedra with fixed orientations.

A central difficulty is that, unlike in the plane, the total volume and the coordinate extents of the objects do not provide a useful lower bound on the optimum: even objects with small total volume and small diameter may require a container of large volume. We overcome this difficulty using a better geometric lower bound based on a surprising connection to Wang and Zahl's celebrated recent proof of the three-dimensional Kakeya conjecture. 

Using this, we obtain a polynomial-time $n^{o(1)}$-approximation algorithm. A recent strengthening of Wang and Zahl's theorem claimed by OpenAI gives a polylogarithmic approximation ratio for the same algorithm.

Our ideas extend to bin and strip packing: for any fixed $\delta>0$, we obtain the same approximation guarantees for translational packing of convex polyhedra whose coordinate widths are at most a $1-\delta$ fraction of the corresponding container widths.
\end{abstract}
\thispagestyle{empty}
\newpage
\setcounter{page}{1}
\section{Introduction}
Packing problems are among the most fundamental problems in discrete and computational geometry. We study the packing of convex polyhedra in three dimensions using translations only. Given $n$ convex polyhedra $K_1,\ldots,K_n$ in $\mathbb{R}^3$, the goal is to translate them into an axis-parallel box of minimum volume so that their interiors are pairwise disjoint.
Boundary contact is allowed.
We denote $\operatorname{OPT}_{\mathrm{vol}}$ for this minimum volume.

Beyond their theoretical appeal, three-dimensional packing problems are of considerable practical importance, with classical applications in container loading, pallet loading, and cutting stock, and more recent applications in additive manufacturing (3D printing). In powder-based technologies such as selective laser sintering, many parts are produced in a single machine run, and because spreading each layer of powder takes much longer than sintering it, the parts should be nested as densely as possible in the build volume~\cite{EgebladNB09}. The resulting layout is a major factor in the material cost and production time of a job~\cite{LamasFernandezBM23}, and the nesting problem has received considerable attention in the manufacturing literature; see the surveys~\cite{OhWLS20,AraujoOAB19}.

The translational case is particularly relevant in additive manufacturing, where the build orientation of each part is usually chosen before nesting because it determines surface quality, the required support structures, and the build time and cost~\cite{GriffithsSEMC19,AraujoPOABA20}. The parts are then placed using translations, possibly combined with rotations about the vertical axis~\cite{OhWLS20,AraujoPOABA20}. Similar restrictions arise in the plane, where the grain or pattern of a material such as fabric may prevent pieces from being rotated~\cite{DBLP:conf/soda/AamandA0K23}; such constraints have motivated the study of translational packing of convex polygons~\cite{DBLP:journals/jocg/AltBK17,DBLP:conf/esa/KurpiszS23,DBLP:conf/soda/AamandA0K23}.

The packing literature is vast; see the surveys~\cite{ChristensenKPT17,AliRCO22,CoffmanCGMV13,WascherHS07,Alt16} with references to hundreds of articles on three-dimensional packing.
Most theoretical work in three dimensions concerns packing axis-parallel cuboids by translations into bins, a strip, or a minimum-volume box, a line of research dating back to the 1960s~\cite{GilmoreG65,MeirM68}. Since the work of Li and Cheng~\cite{LiC90}, approximation algorithms have been developed for three-dimensional bin packing and strip packing~\cite{CsirikV93,MiyazawaW97,EpsteinS05,JansenS06,Caprara08,BansalHISZ13,JansenP14,Sharma21} and for related knapsack and hypercube packing problems~\cite{DiedrichHJTT08,Harren09}.

The emphasis on cuboids reflects what is tractable rather than the range of shapes appearing in practice.
Parts to be printed are rarely cuboids, yet studies of production planning for additive manufacturing often approximate them by rectangles or cuboids to simplify nesting~\cite{OhWLS20}. Practical approaches that handle general polyhedra directly rely on heuristics and mathematical programming models without performance guarantees~\cite{StoyanGSPM05,EgebladNO07,EgebladNB09,LamasFernandezBM23}; see~\cite{LeaoTOCA20} for a review of such models.

\paragraph{Results.}
Since the problem is NP-hard even for axis-parallel cuboids~\cite{DBLP:journals/ijcga/AltS18}, we aim for approximation algorithms and prove the following main result.

\begin{theorem}
\label{thm:main}
There is a polynomial-time $n^{o(1)}$-approximation algorithm for translational packing of $n$ convex polyhedra in $\mathbb{R}^3$ into a minimum-volume axis-parallel box.
\end{theorem}

More precisely, for every $\varepsilon>0$, the approximation ratio is at most $C_{\varepsilon}n^{\varepsilon}$ for some constant $C_{\varepsilon}$, while the algorithm and its running time are independent of $\varepsilon$. The constant $C_{\varepsilon}$ depends on unspecified constants in Wang and Zahl's Kakeya estimate~\cite{WangZahl2025}, so we do not make the $o(1)$ term explicit.

This paper was essentially complete when OpenAI announced a proof of Sogge's local smoothing conjecture~\cite{OpenAILocalSmoothing2026}, accompanied by a Lean formalization~\cite{OpenAILean2026}. The formalization proves a stronger bound than Wang and Zahl's on the volume of unions of direction-separated tubes, while the preprint claims a theorem that would yield an even stronger bound and has not been formally verified. Both make our approximation ratio polylogarithmic, as stated in the following theorem, which is proven in Appendix \ref{sec:polylog}.
 
\begin{theorem}
\label{cor:polylog}
Using OpenAI's formally verified tube estimate~\cite{OpenAILean2026}, the algorithm of \cref{thm:main} achieves approximation ratio $O(\log^{69}n)$. If the Kakeya maximal estimate claimed by OpenAI~\cite[Corollary~9.1]{OpenAILocalSmoothing2026} holds, the ratio improves to $O(\log^7n)$.
\end{theorem}

We consider the minimum-volume box to be the most basic container objective, but the applications also lead to other objectives.
In additive manufacturing, the build volume of the machine is fixed, and one wants to minimize the number of runs needed to produce a given set of parts or, within a single run, the height of the layout, which determines the number of layers and hence the build time~\cite{EgebladNB09,OhWLS20}. These are the three-dimensional versions of two classical problems: in \emph{bin packing}, the objects have to be packed into the minimum number of unit cubes, and in \emph{strip packing}, they have to be packed into a container $[0,1]^2\times[0,h]$ of minimum height $h$. For cuboids, all three objectives have been studied with closely related techniques~\cite{DBLP:journals/ijcga/AltS18,KarKR25}. Our techniques carry over to bin packing and strip packing of convex polyhedra under translations, provided that the extend of the objects is bounded away from that of the container.

\begin{theorem}
\label{thm:bin-strip}
Suppose that translational packing of convex polyhedra in $\mathbb{R}^3$ into a minimum-volume axis-parallel box admits a polynomial-time $\alpha$-approximation. Then, for every $0<\delta<1$, translational bin packing and strip packing of convex polyhedra with coordinate widths at most $1-\delta$ admit polynomial-time $O(\alpha/\delta^3)$- and $O(\alpha/\delta^2)$-approximation algorithms, respectively.
\end{theorem}
We prove \cref{thm:bin-strip} in \cref{sec:bin-strip}. By \cref{thm:main}, we can take $\alpha=n^{o(1)}$, and by \cref{cor:polylog}, even $\alpha=O(\log^{69}n)$.

\paragraph{Previous Techniques.}
Known constant-factor approximation algorithms for packing convex objects rely on what we call the \emph{volume lower bound}, which combines total area or volume with the space needed to accommodate individual objects~\cite{DBLP:journals/jocg/AltBK17,DBLP:journals/ijcga/AltS18}. For translational packing in the plane, for example, this bound is
\begin{equation}
\label{eq:vol-extent}
\operatorname{OPT}_{\mathrm{area}}\geq
\max\left\{\sum_i\operatorname{area}(K_i),\;
w_xw_y\right\},
\end{equation}
where $\operatorname{OPT}_{\mathrm{area}}$ is the minimum area of an axis-parallel rectangle containing a translational packing, and $w_x,w_y$ are the maximum widths of the input polygons in the respective coordinate directions.

In the plane, even under translations, Alt, de Berg, and Knauer~\cite{DBLP:journals/jocg/AltBK17,DBLP:journals/jocg/BergK20} gave constant-factor approximations for packing convex polygons, and Kurpisz and Suter~\cite{DBLP:conf/esa/KurpiszS23} improved the factor for minimum-area rectangles to $9.45$. Their analysis shows that the volume lower bound~\eqref{eq:vol-extent} suffices for a constant-factor approximation.

Analogous versions of the volume lower bound are particularly useful under rigid motions, where rotations are allowed: they yield constant-factor approximations for packing cuboids and general convex objects~\cite{DBLP:journals/jocg/AltBK17,DBLP:journals/ijcga/AltS18}. The same approach extends to any fixed dimension $d$, with constants depending on $d$.

Whether a constant-factor approximation exists for translational packing of arbitrary convex polyhedra in three dimensions has repeatedly been posed as an open question~\cite{Seiferth22,DBLP:journals/comgeo/AltCCPS26,AamandABKKT26}. Theorem~\ref{thm:main} gives the first approximation guarantee for this problem.
Table~\ref{tab:related-work} summarises the known results on minimum-area and minimum-volume packing.

For fixed orientations in three dimensions, the volume/max-extent bound~\eqref{eq:vol-extent} can be far too small. Consider $n$ unit-radius planar disks in $\mathbb R^3$, with arbitrary normal vectors: their total three-dimensional volume is zero and each coordinate width is at most $2$, so the bound is at most $8$. Yet there are choices of normal vectors for which every translational packing requires volume $\Omega(\sqrt n)$~\cite{DBLP:journals/comgeo/AltCCPS26}. Enclosing these disks in sufficiently thin convex polyhedra gives the same obstruction with full-dimensional objects.
One of the missing ingredients to obtain packing algorithms in three dimensions is therefore a stronger lower bound on the cost of the optimal solution that captures the space consumption forced by such different orientations. The second ingredient is an algorithm that comes close to this new lower bound.

For unit disks, Alt, Cabello, Cheong, Park, and Seiferth~\cite{DBLP:journals/comgeo/AltCCPS26} obtain such a lower bound through a Hamiltonian-path formulation.
They prove that in any packing, the disks can be rearranged into stacks, with the disk centers on lines in the directions of the three coordinate axis.
Placing the disks like that defines a metric, which is the distance between centers.
It follows that arranging the disks according to a constant factor approximation of the shortest Hamiltonian path yields a constant-factor approximation algorithm.
The argument relies on equal radii, and the authors state the case of disks having different radii as an open problem. They extend the algorithm for unit disks to congruent copies of a single planar shape, however this loses a factor of $O(\rho^3)$, where $\rho$ is the ratio of the radii of an enclosing and inscribed disk of the shape. Thus this gives no uniform guarantee for arbitrary polyhedra.

\begin{table}[h]
    \centering
    \small
    \renewcommand{\arraystretch}{1.25}
    \begin{tabularx}{\textwidth}{@{}lXX@{}}
        \toprule
        & \textbf{Rigid motions} & \textbf{Translations} \\
        \midrule
        \textbf{2D}
        & Convex polygons: $O(1)$~\cite{DBLP:journals/jocg/AltBK17}
        & Convex polygons: $9.45$~\cite{DBLP:conf/esa/KurpiszS23} \\
        \midrule
        \textbf{3D}
        & Convex polyhedra: $O(1)$~\cite{DBLP:journals/ijcga/AltS18}
        & Axis-parallel cuboids: $O(1)$~\cite{DBLP:journals/ijcga/AltS18}, $3+\varepsilon$ and APTAS~\cite{KarKR25}\par\smallskip
          Unit disks (planar) in $\mathbb R^3$: $O(1)$~\cite{DBLP:journals/comgeo/AltCCPS26}\par\smallskip
          Congruent planar shapes: $O(\rho^3)$~\cite{DBLP:journals/comgeo/AltCCPS26}\par\smallskip
          Convex polyhedra: $\boldsymbol{n^{o(1)}}$, $O(\log^{69}n)$, $O(\log^7n)$ \textbf{(this work)} \\
        \bottomrule
    \end{tabularx}
    \caption{Approximation guarantees for minimum-area (2D) and minimum-volume
    (3D) axis-parallel containers.
    Rigid motions allow rotations. For congruent planar shapes, $\rho$
    is the ratio of enclosing to inscribed disk radii.}
    \label{tab:related-work}
\end{table}

\subsection{The Kakeya volume estimate and its relation to packing.}
\label{sec:preliminaries}

Our geometric input is the following consequence of Wang and Zahl's theorem~\cite{WangZahl2025}. For $v\in S^2$, $x\in\mathbb R^3$, and $\delta>0$, the \emph{tube} $T_\delta(v,x)$ is the open $\delta$-neighbourhood of the unit segment $\{x+tv:|t|\leq1/2\}$. For $\delta\leq1$, it has volume $\pi\delta^2+\frac43\pi\delta^3\leq\frac73\pi\delta^2$.
For vectors $v,w\in \mathbb{R}^3\setminus \{0\}$, we define the \emph{unoriented angle} between them to be the minimum of the angle between $v_i$ and $v_j$ and the angle between $v_i$ and $-v_j$.
\begin{theorem}
\label{thm:tube-union}
For every $\varepsilon>0$, there exists a constant $C_\varepsilon\geq1$ such that the following holds for all $0<\delta\leq1$. Let $v_1,\ldots,v_m\in S^2$ be such that for distinct $i,j$, the
unoriented angle between $v_i$ and $v_j$ is at least $\delta$, and let $x_1,\ldots,x_m\in\mathbb R^3$. Then
\[
\operatorname{vol}\!\left(\bigcup_{i=1}^mT_\delta(v_i,x_i)\right)
\geq\frac{\delta^\varepsilon}{C_\varepsilon}
\sum_{i=1}^m\operatorname{vol}(T_\delta(v_i,x_i)).
\]
\end{theorem}

In Appendix \ref{app:tubes} we prove that Theorem~\ref{thm:tube-union} follows from~\cite[Theorem~1.2]{WangZahl2025}.

\paragraph{Kakeya lower bounds.}

We briefly discuss why the Kakeya volume estimate is useful for providing lower bounds for packings. Our main concern is flat pieces where a volume bound is insufficient. For a packing of such flat pieces, it turns out to be a very useful technique to identify a subset of the pieces such that after thickening them to some thickness $\tau$, there is a linear transformation that maps the pieces to approximate Kakeya tubes. Then the Kakeya volume estimate tells us that there cannot be too much overlap between the thickened pieces, and the volume occupied by the thickened pieces can be lower bounded in terms of the sums of their volumes. Within constant factors, the cost of an optimal solution is at least that of the volume occupied by thickened pieces, so this gives a lower bound of its cost in terms of the number of such pieces. We discuss this in a bit more detail below.

\subsection{Discussion of our techniques}
We enclose the polyhedra in parallelepipeds and round two edge parameters to obtain $O(\log^2 n)$ types, with only a constant-factor enlargement. Within each type, the shadows (projections) onto the $yz$-plane all have the same area $A$, and their shadows fall into a discrete set of possible parallelograms. In the following, we explain how we handle a single type.

\paragraph{Upper bound.}
We stack objects with the same shadow by placing their centres on a line parallel to the $x$-axis so that their shadows coincide. The space needed depends on how closely consecutive objects fit together, and this separation cost defines a metric on each shadow group. Choosing an order with minimum total separation is therefore a metric Hamiltonian-path problem. We use the classic MST-based $2$-approximation to order each group and then combine the stacks. Writing $\operatorname{MST}_k$ for the tree weight of shadow group $k$, the resulting packing has volume $O\!\left(\operatorname{OPT}_{\mathrm{vol}}+A\sum_k\operatorname{MST}_k\right)$, where $\operatorname{OPT}_{\mathrm{vol}}$ is the optimal cost for packing this type of pieces.

\paragraph{Lower bound.}
We need a lower bound on the optimum in terms of the same spanning-tree weights. For intuition, imagine that all tree-edge lengths belong to $\{\frac{1}{n},\frac{2}{n},\frac{4}{n},\ldots,1\}$, giving only $O(\log n)$ possible values for the scale. Some value, say $\tau_1$, accounts for $\Omega\!\left((\sum_k\operatorname{MST}_k)/\log n\right)$ of the total weight, so there are $\Omega\!\left((\sum_k\operatorname{MST}_k)/(\tau_1\log n)\right)$ edges of that length. These edges of the MSTs allow us to select $m=\Omega\!\left((\sum_k\operatorname{MST}_k)/(\tau_1\log n)\right)$ representative objects such that any two from the same shadow group have distance at least $\tau_1$. Indeed, by the MST property, after removing the edges of length at least $\tau$, we end up with connected components such that the distance between any two pieces in different connected components is at least $\tau$, so we can select a representative from each such component.

Start with any feasible packing (in particular, an optimal one) and thicken each representative in the $x$-direction to thickness $\Theta(\tau_1)$. Each resulting object has volume $\Theta(A\tau_1)$, so their total volume is $\Omega\!\left(\frac{A\sum_k\operatorname{MST}_k}{\log n}\right)$. If they were disjoint, their union would already give the desired lower bound, up to this logarithmic factor. 
They may overlap, but we argue that we can select a large subset and partition it into groups so that objects from different groups have disjoint interiors. Within each group, we apply a common linear transformation that stretches and shears the objects so that they roughly correspond to tubes in the transformed space. Intuitively, the tube directions are separated because any two objects either belong to different shadow classes or have distance at least $\tau_1$ within the same class. We can now apply \cref{thm:tube-union} within each group to show that there is little overlap, then transfer the volume bounds back to the original space and add them across the disjoint groups.
Making this possible while retaining enough objects is our main technical contribution. The union lies in a constant enlargement of the container, so the Kakeya estimate gives, for every $\varepsilon>0$,
\[
A\sum_k\operatorname{MST}_k=O_\varepsilon(n^\varepsilon)\operatorname{OPT}_{\mathrm{vol}}.
\]
Together with the upper bound, this gives the approximation guarantee. Stronger tube estimates give the polylogarithmic guarantees for the same algorithm.

\section{Packing \texorpdfstring{$bc$}{bc}-parallelepipeds}
\label{sec:algorithm}

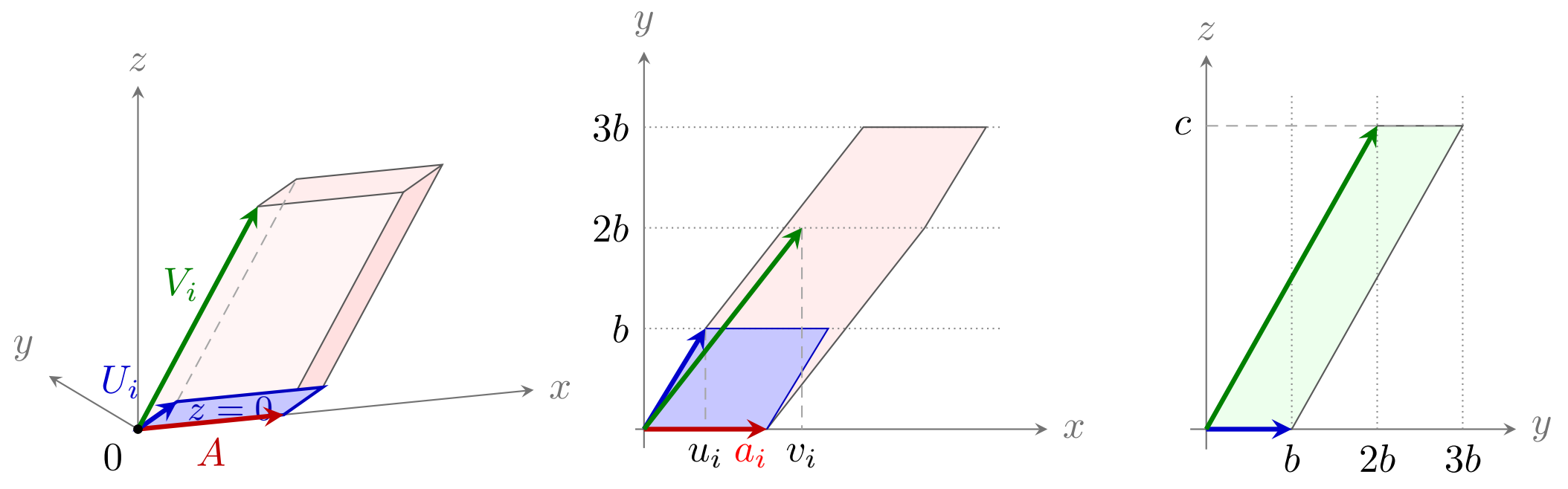
\begin{figure}[h]
    \centering
    \begin{tikzpicture}[
        >=stealth,
        every node/.style={font=\small},
        edge/.style={draw=black!65,thin},
        first/.style={red!75!black,very thick,->},
        second/.style={blue!80!black,very thick,->},
        third/.style={green!50!black,very thick,->},
        guide/.style={black!35,densely dashed,thin},
        axis/.style={->,black!55,thin}
    ]
        \begin{scope}[xshift=1.1cm]
            \begin{scope}[x={(.9262cm,.0919cm)},y={(-.3770cm,.2256cm)},z={(0cm,.9699cm)}]
                \coordinate (O) at (0,0,0);
                \coordinate (A) at (1.4,0,0);
                \coordinate (U) at (.7,.8,0);
                \coordinate (V) at (1.8,1.6,1.5);
                \coordinate (AU) at (2.1,.8,0);
                \coordinate (AV) at (3.2,1.6,1.5);
                \coordinate (UV) at (2.5,2.4,1.5);
                \coordinate (AUV) at (3.9,2.4,1.5);
                \coordinate (base-center) at (1.05,.4,0);
                \draw[axis] (0,0,0)--(3.8,0,0) node[right] {$x$};
                \draw[axis] (0,0,0)--(0,2.1,0) node[above left] {$y$};
                \draw[axis] (0,0,0)--(0,0,3.15) node[above] {$z$};
            \end{scope}
            \fill[red!4] (O)--(A)--(AV)--(V)--cycle;
            \fill[red!12] (A)--(AU)--(AUV)--(AV)--cycle;
            \fill[red!7] (V)--(AV)--(AUV)--(UV)--cycle;
            \draw[edge] (O)--(A)--(AU)--(AUV)--(UV)--(V)--cycle;
            \draw[edge] (A)--(AV)--(V) (AV)--(AUV);
            \draw[guide] (O)--(U)--(AU) (U)--(UV);
            \filldraw[fill=blue!22,draw=blue!75!black,thick] (O)--(A)--(AU)--(U)--cycle;
            \node[font=\footnotesize,text=blue!70!black] at (base-center) {$z=0$};
            \draw[first] (O)--(A) node[midway,below] {$A$};
            \draw[second] (O)--(U) node[midway,above left] {$U_i$};
            \draw[third] (O)--(V) node[pos=.65,left] {$V_i$};
            \fill (O) circle[radius=1.2pt];
            \node[below left] at (O) {$0$};
            \node at (1.05,-1.0) {(a) $bc$-parallelepiped};
        \end{scope}

        \begin{scope}[xshift=5.6cm,x=.78cm,y=1.12cm]
            \draw[axis] (-.1,0)--(4.6,0) node[right] {$x$};
            \draw[axis] (0,-.15)--(0,3.0) node[above] {$y$};
            \fill[red!7] (0,0)--(1.4,0)--(3.2,1.6)--(3.9,2.4)--(2.5,2.4)--(.7,.8)--cycle;
            \fill[blue!22] (0,0)--(1.4,0)--(2.1,.8)--(.7,.8)--cycle;
            \foreach \level/\lab in {.8/b,1.6/2b,2.4/3b} {
                \draw[black!45,densely dotted] (0,\level)--(4.1,\level);
                \node[left] at (0,\level) {$\lab$};
            }
            \draw[edge] (0,0)--(1.4,0)--(3.2,1.6)--(3.9,2.4)--(2.5,2.4)--(.7,.8)--cycle;
            \draw[blue!75!black,thin] (1.4,0)--(2.1,.8)--(.7,.8);
            \draw[guide] (.7,.8)--(.7,0) (1.8,1.6)--(1.8,0);
            \node[below] at (.7,0) {$u_i$};
            \node[below] at (1.8,0) {$v_i$};
            \draw[first] (0,0)--(1.4,0);
            \draw[second] (0,0)--(.7,.8);
            \draw[third] (0,0)--(1.8,1.6);
            \node[below,text=red,xshift=-4pt] at (1.4,0) {$a_i$};
            \node at (2.05,-.8929) {(b) Projection onto $xy$};
        \end{scope}

        \begin{scope}[xshift=10.6cm,x=.95cm,y=1.8cm]
            \draw[axis] (-.15,0)--(2.9,0) node[right] {$y$};
            \draw[axis] (0,-.1)--(0,1.85) node[above] {$z$};
            \fill[green!7] (0,0)--(.8,0)--(2.4,1.5)--(1.6,1.5)--cycle;
            \foreach \level/\lab in {.8/b,1.6/2b,2.4/3b} {
                \draw[black!45,densely dotted] (\level,0)--(\level,1.65);
                \node[below] at (\level,0) {$\lab$};
            }
            \draw[edge] (0,0)--(.8,0)--(2.4,1.5)--(1.6,1.5)--cycle;
            \draw[guide] (0,1.5)--(2.4,1.5);
            \node[left] at (0,1.5) {$c$};
            \draw[second] (0,0)--(.8,0);
            \draw[third] (0,0)--(1.6,1.5);
            \node at (1.25,-.5556) {(c) Projection onto $yz$};
        \end{scope}
    \end{tikzpicture}
    \caption{A $bc$-parallelepiped with $k_i=2$, translated so that one vertex is at the origin (a), and its projections onto the $xy$-plane (b) and the $yz$-plane (c). The edges are $A=(a_i,0,0)$, $U_i=(u_i,b,0)$, and $V_i=(v_i,k_i b,c)$. Objects with the same $k_i$ share the $yz$-projection.}
    \label{fig:bc-parallelepiped}
\end{figure}

We first describe the packing algorithm for a special family of
parallelepipeds.~\Cref{sec:reduction} reduces general convex polyhedra to
such families. Each object has one edge parallel to the $x$-axis, one edge
parallel to the $xy$-plane, and one edge with a positive $z$-component (see
\cref{fig:bc-parallelepiped}). 

We fix the $y$-component $b$ of the second edge, and the $z$-component $c$ of the third edge.
The components $u_i,v_i$, and $k_i$ of the second and third edges describe how the objects slant.

More precisely, for fixed $b,c>0$, a \emph{$bc$-parallelepiped} is a
translate of 
\[
P_i=\left\{
\lambda_1\begin{pmatrix}a_i\\0\\0\end{pmatrix}
+\lambda_2\begin{pmatrix}u_i\\b\\0\end{pmatrix}
+\lambda_3\begin{pmatrix}v_i\\k_i b\\c\end{pmatrix}
:\lambda_1,\lambda_2,\lambda_3\in[-\tfrac12,\tfrac12]
\right\},
\]
where $a_i >0$, $u_i,v_i\in\mathbb R$ and $k_i\in\mathbb Z$ are the four free parameters. Here $a_i$ is the length of the first edge, $u_i$ and $v_i$ are the $x$-components of the second and third edges, while $k_i$ specifies the third edge's $y$-component in multiples of $b$.
Objects sharing $(b,c)$ form a \emph{type}, and object $i$ has volume $a_ibc$.
We consider a single type with $n$ objects $P_1,\ldots,P_n$. We assume that each object fits individually in the unit cube, or equivalently that
\[
a_i+|u_i|+|v_i|\leq1,\qquad(1+|k_i|)b\leq1,\qquad c\leq1.
\]

The projection onto the $yz$-plane maps the three edges to $(0,0)$, $(b,0)$, and $(k_ib,c)$, so the shadow of $P_i$ is the parallelogram spanned by $(b,0)$ and $(k_ib,c)$; see \cref{fig:bc-parallelepiped}(c). Within a type, this shadow depends only on $k_i$, not on $u_i$ and $v_i$. Objects with $k_i=k$ form \emph{shadow class} $k$.

One way to pack objects of the same shadow class is to translate them only along the $x$-axis, so that their shadows coincide. The distance required between two such objects is determined by their \textit{shears} $u_i$ and $v_i$. For two objects $i$ and $j$, let
\[
d(i,j)=|u_i-u_j|+|v_i-v_j|.
\]
The following lemma shows that their centres then need to be exactly $\frac{a_i+a_j+d(i,j)}{2}$ apart.

\begin{lemma}
\label{lem:shadow-class}
Let $i$ and $j$ be objects of the same shadow class. Then $P_i$ and $P_j+(h,0,0)$ have disjoint interiors if and only if $|h|\geq\frac{a_i+a_j}2+\frac{d(i,j)}2$.
\end{lemma}

\begin{proof}
The common shadow of the shadow class consists of the points $(bs+kbt,ct)$ with $s,t\in[-\frac12,\frac12]$. The point $\lambda_1(a_i,0,0)+\lambda_2(u_i,b,0)+\lambda_3(v_i,k_ib,c)$ of $P_i$ lies above $(bs+kbt,ct)$ exactly when $\lambda_2=s$ and $\lambda_3=t$. Hence, above this point, $P_i$ and $P_j+(h,0,0)$ occupy the $x$-intervals $u_is+v_it+[-\frac{a_i}2,\frac{a_i}2]$ and $h+u_js+v_jt+[-\frac{a_j}2,\frac{a_j}2]$. These intervals overlap in an interval of positive length exactly when the distance between their midpoints is less than $\frac{a_i+a_j}2$. The distance between their midpoints is $|\delta_{ij}(s,t)|$, where
\[
\delta_{ij}(s,t)=h+(u_j-u_i)s+(v_j-v_i)t.
\]
As $s$ and $t$ range over $[-\frac12,\frac12]$, the values of $\delta_{ij}$ form the interval $[h-\frac{d(i,j)}2,\,h+\frac{d(i,j)}2]$, with the endpoints attained at corners.

If $|h|\geq\frac{a_i+a_j}2+\frac{d(i,j)}2$, then either $h-\frac{d(i,j)}2\geq\frac{a_i+a_j}2$ or $h+\frac{d(i,j)}2\leq-\frac{a_i+a_j}2$, so $|\delta_{ij}|\geq\frac{a_i+a_j}2$ everywhere. Hence, above every point of the shadow, the two intervals share at most an endpoint, so the interiors of the objects do not overlap.

Conversely, let $|h|<\frac{a_i+a_j}2+\frac{d(i,j)}2$. The value of the interval closest to $0$ has absolute value $\max\{|h|-\frac{d(i,j)}2,0\}<\frac{a_i+a_j}2$, so $|\delta_{ij}(s,t)|<\frac{a_i+a_j}2$ for some $s,t$ with $|s|,|t|<\frac12$. Above the corresponding interior point of the shadow, the two $x$-intervals overlap in an interval of positive length, whose interior points are interior points of both objects.
\end{proof}

Thus, placing objects with similar $u_i,v_i$-values consecutively saves space. We write $\operatorname{MST}_k$ for the minimum spanning tree weight of shadow class $k$ under $d$.

\subsection{The algorithm}
The algorithm places the objects in three steps, illustrated in \cref{fig:algorithm-fan-3d} and summarized in \cref{alg:canonical-packing}.
\input{figures/algorithm-fan-3d}

\begin{enumerate}[(1)]
\item \emph{Order each shadow class.} We order the class by the classic MST-based $2$-approximation for metric Hamiltonian paths; consecutive objects in this order have total distance at most $2\operatorname{MST}_k$. We translate object $i$ by $(x_i,0,0)$, where the first object gets the offset $x_i=0$, and every later object gets the offset $x_j=x_i+\frac{a_i+a_j}2+\frac{d(i,j)}2$, where $i$ is its predecessor, so consecutive objects then touch without overlapping.
\item \emph{Cut the classes into blocks.} Objects of the same class with the same value of $\lfloor x_i\rfloor$ form a \emph{block}, and we translate each block by $(-\lfloor x_i\rfloor,0,0)$, so that all offsets lie in $[0,1)$. Since $a_i+|u_i|+|v_i|\leq1$, every object lies within $x$-distance $1/2$ of its offset, so every block lies in the slab $-1/2\leq x\leq3/2$, and no object is cut. Without this step, a single long shadow class could determine the $x$-extent of the whole packing. For a visualization, compare the two subfigures of \cref{fig:algorithm-fan-3d}, where on the right, the turquoise shadow class has been cut into blocks.
\item \emph{Place the blocks in a fan.} We number the blocks $1,\ldots,N$ in order of increasing shadow class $k$, breaking ties arbitrarily, and translate the $\ell$-th block, of class $k$, by $(0,(\ell-1+k/2)b,0)$, so that the bottom edges of the block shadows lie next to each other. Since the slants increase from left to right, the shadows spread apart as $z$ increases and never cross.
\end{enumerate}

\begin{algorithm}[h]
\caption{Packing $bc$-parallelepipeds}
\label{alg:canonical-packing}
\begin{algorithmic}[1]
\For{each nonempty shadow class $k$}
    \State Order the class as $i_1,\ldots,i_r$ by the MST-based $2$-approximation.
    \State Set $x_{i_1}\gets0$ and $x_{i_j}\gets x_{i_{j-1}}+\frac{a_{i_{j-1}}+a_{i_j}}2+\frac{d(i_{j-1},i_j)}2$ for $j=2,\ldots,r$.
\EndFor
\State Number the blocks, i.e.\ the distinct pairs $(k_i,\lfloor x_i\rfloor)$, as $1,\ldots,N$ in order of increasing $k_i$.
\State Let $\ell_i$ be the number of the block containing $P_i$.
\State Translate each $P_i$ by $(x_i-\lfloor x_i\rfloor,\,(\ell_i-1+k_i/2)b,\,0)$.
\State \Return these translations and the axis-parallel bounding box.
\end{algorithmic}
\end{algorithm}

The next lemma ensures that objects in different blocks do not overlap.

\begin{lemma}
\label{lem:fan-disjoint}
Let $P_i$ and $P_j$ be objects with $k_i\leq k_j$, and let $\ell_i<\ell_j$ be integers. Then, for all $x_i,x_j\in\mathbb R$, the objects $P_i+(x_i,(\ell_i-1+k_i/2)b,0)$ and $P_j+(x_j,(\ell_j-1+k_j/2)b,0)$ have disjoint interiors.
\end{lemma}

\begin{proof}
At height $z=ct$, where $-\frac12\leq t\leq\frac12$, the two objects have $y$-coordinates in $(\ell_i-1)b+k_ib(t+\frac12)+[-\frac b2,\frac b2]$ and $(\ell_j-1)b+k_jb(t+\frac12)+[-\frac b2,\frac b2]$, respectively. The second interval starts where the first ends or later, because the difference of their midpoints is
$(\ell_j-\ell_i)b+(k_j-k_i)b\bigl(t+\tfrac12\bigr)\geq b$.

\end{proof}

\begin{lemma}
\label{lem:fan}
In $O(n\log n)$ time, \cref{alg:canonical-packing} returns a translational packing of $P_1,\ldots,P_n$ in an axis-parallel box of volume
\[
O\!\left(\sum_i a_ibc+bc\sum_k\operatorname{MST}_k+c\right).
\]
\end{lemma}

\begin{proof}
We first show that no two objects overlap. Consider two objects $i\neq j$, labelled so that $\ell_i\leq\ell_j$. If they lie in the same block, their translations differ only in the $x$-coordinate, by $x_j-x_i$, and we may assume that $i$ comes before $j$ in their class. The distance $d$ is the $\ell_1$-distance between the points $(u_i,v_i)$, so the triangle inequality gives $x_j-x_i\geq\frac{a_i+a_j}2+\frac{d(i,j)}2$, and \cref{lem:shadow-class} applies. If they lie in different blocks, then $\ell_i<\ell_j$ and hence $k_i\leq k_j$, so \cref{lem:fan-disjoint} applies.

Next, we bound the volume. The offsets of class $k$ lie in $[0,A_k+\operatorname{MST}_k]$, where $A_k$ is the sum of the $a_i$ in the class, so the class produces at most $A_k+\operatorname{MST}_k+1$ blocks. Since $|k|b\leq1$, at most $1+2/b$ classes are nonempty, so the total number $N$ of blocks satisfies
\[
Nb\leq\sum_ia_ib+b\sum_k\operatorname{MST}_k+b\Bigl(1+\frac2b\Bigr)
\leq3+\sum_ia_ib+b\sum_k\operatorname{MST}_k.
\]
The final packing lies in $[-\frac12,\frac32]\times[-1-\frac b2,(N-\frac12)b+1]\times[-\frac c2,\frac c2]$, since the $\ell$-th block has $y$-coordinates within $(1+|k|)b/2$ of $(\ell-1+k/2)b$ and $|k|b\leq1$. Multiplying the side lengths proves the bound.

Finally, we bound the running time. The distance $d$ is the $\ell_1$-distance between the points $(u_i,v_i)$, so a minimum spanning tree of a class with $r$ objects can be computed in $O(r\log r)$ time~\cite{Hwang79}. Finally, sorting the at most $n$ blocks by shadow class takes $O(n\log n)$ time.
\end{proof}

\subsection{Bounding the spanning-tree cost}
\label{sec:spanning-tree-term}

Since $\sum_ia_ibc\leq1$ and $c\leq1$ for families that fit in the unit cube, it remains to bound $bc\sum_k\operatorname{MST}_k$.

For a shadow class $k$, let $G_k$ be the complete graph on its objects with edge weights $d(i,j)$. For $\tau>0$, define
\[
G_k(\tau)=\bigl(V(G_k),\{e\in E(G_k):d(e)\leq\tau\}\bigr),
\]
where $d(e)$ is the weight of edge $e$, and let $c_k(\tau)$ be its number of connected components. Since each object fits in the unit cube, $|u_i|+|v_i|\leq1$, so all edge weights are at most $2$. We have the identity
\begin{equation}
\label{eq:mst-identity}
\operatorname{MST}_k
=
\int_0^2 \bigl(c_k(\tau)-1\bigr)\,d\tau.
\end{equation}

Indeed, a minimum spanning tree has exactly $c_k(\tau)-1$ edges of weight greater than $\tau$: deleting these edges leaves a forest whose trees exactly correspond to the components of $G_k(\tau)$. Each tree lies in one component, since its edges belong to $G_k(\tau)$.
Conversely, if there are two or more trees in the same connected component, then there is also an edge $xy$ of $G_k(\tau)$ joining two such trees.
Hence, the path from $x$ to $y$ in the minimum spanning tree would contain a deleted edge, which is heavier than $xy$. Exchanging the two edges would give a lighter spanning tree.
An edge of weight $w$ of the minimum spanning tree is therefore counted for $0\leq\tau<w$ and contributes exactly $w$ to the integral. This is a continuous version of an identity of Chazelle, Rubinfeld, and Trevisan~\cite[Claim~5]{ChazelleRT05}.

Summing over all shadow classes and exchanging sum and integral gives
\[
\sum_k \operatorname{MST}_k
=
\int_0^2 \sum_k \bigl(c_k(\tau)-1\bigr)\,d\tau.
\]
The point of this identity is that it replaces the spanning trees, which mix all scales, by component counts at single scales. At a fixed scale, many components mean many objects whose shears are pairwise far apart, and a packing in the unit cube cannot contain too many such objects, as we show below. It therefore suffices to bound $\sum_k c_k(\tau)$, the total number of components at scale~$\tau$, for every~$\tau$. We bound this sum for all shadow classes together, since bounding each class separately would lose a factor of the number of classes.

Fix $\tau$ and choose one representative from each component of every $G_k(\tau)$. These $\sum_k c_k(\tau)$ objects still fit in the unit cube, and two representatives of the same shadow class satisfy $d(i,j)>\tau$, since otherwise an edge would join their components. The following theorem, the geometric lower bound at the core of our analysis, therefore bounds $\sum_k c_k(\tau)$; we prove it in \cref{sec:lower-bound}.

\begin{theorem}[Lower Bound]
\label{thm:counting}
For every $\varepsilon>0$, there exists a constant $C_\varepsilon>0$ such that the following holds. Let $\mathcal P$ be a family of at most $n$ $bc$-parallelepipeds that can be packed by translations into the unit cube. If, for some $0<\tau\leq 2$, we have $d(P_i, P_j)>\tau$ for every pair of distinct objects $P_i,P_j\in\mathcal P$ with $k_i=k_j$, then
\[
|\mathcal P|
\leq C_\varepsilon\,\frac{n^{\varepsilon}}{bc\tau}.
\]
\end{theorem}

Note that the theorem holds for arbitrary separated families and is not specific to the representatives. They do, however, satisfy its assumptions, so we can apply it to them in particular:
\begin{enumerate}[(i)]
\item they are $bc$-parallelepipeds of a single type;
\item they form a subfamily of the input, so they can be packed by translations into the unit cube;
\item any two representatives with $k_i=k_j$ lie in different components of $G_{k_i}(\tau)$, so $d(P_i,P_j)>\tau$.
\end{enumerate}
Hence, for every $0<\tau\leq2$,
\[
\sum_k c_k(\tau)
\leq C_\varepsilon\,\frac{n^{\varepsilon}}{bc\tau}.
\]

It remains to integrate over $\tau$ from 0 to 2. Small scales $\tau<1/n$ are handled by the trivial bound of $n$ components, and larger scales by \cref{thm:counting}.

\begin{lemma}
\label{lem:mst}
If $P_1,\ldots,P_n$ are $bc$-parallelepipeds that admit a translational packing into the unit cube, then, for every $\varepsilon>0$,
\[
bc\sum_k\operatorname{MST}_k=O_\varepsilon(n^\varepsilon).
\]
\end{lemma}

\begin{proof}
Choosing one object from each component of the graphs $G_k(\tau)$ gives a family satisfying \cref{thm:counting}, so
$
\sum_k c_k(\tau)\leq
\min\left\{n,\frac{C_\varepsilon n^\varepsilon}{bc\tau}\right\}.
$
The MST identity~\eqref{eq:mst-identity}, split at $1/n$, now gives
\[
\begin{aligned}
bc\sum_k\operatorname{MST}_k
&=bc\int_0^2\sum_k(c_k(\tau)-1)\,d\tau\\
&\leq bc\int_0^{1/n}\sum_k c_k(\tau)\,d\tau+bc\int_{1/n}^2\sum_k c_k(\tau)\,d\tau\\
&\leq bc\int_0^{1/n}n\,d\tau+\int_{1/n}^2\frac{C_\varepsilon n^\varepsilon}{\tau}\,d\tau\\
&=bc+C_\varepsilon n^\varepsilon\ln(2n)\\
&\leq1+C_\varepsilon n^\varepsilon\ln(2n).
\end{aligned}
\]
Finally, we absorb the logarithm and rescale $\varepsilon$.
\end{proof}

Together with \cref{lem:fan}, and since $\sum_ia_ibc\leq1$ and $c\leq1$, this shows that if $P_1,\ldots,P_n$ admit a translational packing into the unit cube, then \cref{alg:canonical-packing} packs them into a box of volume $O_\varepsilon(n^\varepsilon)$. In \cref{sec:reduction}, we scale a general instance so that an optimal container becomes the unit cube and reduce it to this case at the cost of constant and polylogarithmic factors, which gives the approximation ratio $n^{o(1)}$ of \cref{thm:main}.

\section{The lower bound}
\label{sec:lower-bound}

In this section, we prove \cref{thm:counting} by a purely geometric argument.

We fix a family $\mathcal P$ satisfying its assumptions, so that $d(i,j)>\tau$ for any two distinct objects $i,j$ with $k_i=k_j$, and fix a packing in $[0,1]^3$. Throughout the rest of this section, $P_i$ denotes the placed object and $p_i$ its center. Its parameters satisfy
\[
a_i+|u_i|+|v_i|\leq1,\qquad (1+|k_i|)b\leq1,\qquad c\leq1.
\]

Object $i$ has volume $a_ibc$, so volume counting shows that at most $100/(bc\tau)$ objects have $a_i\geq\tau/100$. For objects with $a_i\ll\tau$, however, volume counting is far too weak. We therefore thicken these objects in the $x$-direction so that each has volume roughly $bc\tau$, as needed for the desired counting bound. Thickening may introduce overlaps, but if we can keep these small, the volume of the union will still bring us close to that bound.

We first divide the objects into \textit{slope groups} and construct thickened bodies that can overlap only within the same group, while retaining a constant fraction of the objects. Within each group, the bodies have similar shapes but point in different directions. We apply a common linear transformation so that they contain comparable tubes, and use Kakeya to bound their union volume in the transformed space. Dividing by the absolute determinant of the transformation then gives a volume bound in the original space.
Since bodies from different groups are disjoint, these  bounds add to give the desired counting estimate.

\paragraph{Auxiliary bodies}

By volume counting, at most $100/(bc\tau)$ objects have $a_i\geq\tau/100$, so it suffices to bound the number of objects with $a_i<\tau/100$, and we consider only these from now on; let $\mathcal P'$ be the family of these objects. We cut the real line into intervals of length $\tau/100$ and label them $1,2,\ldots,10,1,2,\ldots,10,\ldots$ from left to right.
\begin{figure}[h]
    \centering
    \begin{tikzpicture}[>=stealth,every node/.style={font=\small}]
        \foreach \col/\j in {blue/2,green!55!black/12,red/22} {
            \fill[\col!14] ({.42*\j},-.19) rectangle ({.42*(\j+1)},.19);
            \draw[\col!75!black,line width=1pt]
                ({.42*\j},0)--({.42*(\j+1)},0);
        }
        \draw[black!65] (-.12,0)--(10.64,0);
        \foreach \j in {0,...,25}
            \draw[black!80] ({.42*\j},-.12)--({.42*\j},.12);
        \foreach \j in {0,...,24} {
            \pgfmathtruncatemacro{\intervalLabel}{mod(\j,10)+1}
            \node[font=\scriptsize,text=black!80] at ({.42*(\j+.5)},.35)
                {$\intervalLabel$};
        }
        \foreach \col/\j in {blue/2,green!55!black/12,red/22}
            \node[font=\scriptsize,text=\col!80!black] at ({.42*(\j+.5)},.35) {$3$};
        \node[right] at (10.73,0) {$\mathbb R$};
        \foreach \j in {0,1,3,4,5,6,7,8,9,10,11,13,14,15,16,17,18,19,20,21,23,24}
            \fill[black!55] ({.42*(\j+.35+.25*mod(\j,2))},0) circle[radius=1.6pt];
        \foreach \col/\j in {blue/2,green!55!black/12,red/22}
            \foreach \s in {.23,.51,.78}
                \fill[\col!80!black] ({.42*(\j+\s)},0) circle[radius=1.5pt];
        \draw[black!65,thin,densely dashed] (7.98,-.2)--(7.98,-.67) (8.4,-.2)--(8.4,-.67);
        \draw[<->,black!85] (7.98,-.57)--(8.4,-.57)
            node[midway,below] {$\tau/100$};
        \draw[black!65,thin,densely dashed] (1.26,-.2)--(1.26,-.67) (5.04,-.2)--(5.04,-.67);
        \draw[<->,black!85] (1.26,-.57)--(5.04,-.57)
            node[midway,below] {$9\tau/100$};
        \node[text=green!.30!black] at (5.25,1.03) {retain label $3$};
        \draw[->,green!45!black] (5.25,.81)--(5.25,.55);
        \draw[red!65!black,thin] (9.45,-.06)--(9.45,-.34)
            node[below] {$u_i$};
    \end{tikzpicture}
    \caption{Partitioning the offsets $u_i$ into slope groups.
    Each point represents one object; the highlighted intervals
    illustrate retaining label $3$.}
    \label{fig:interval-groups}
\end{figure}
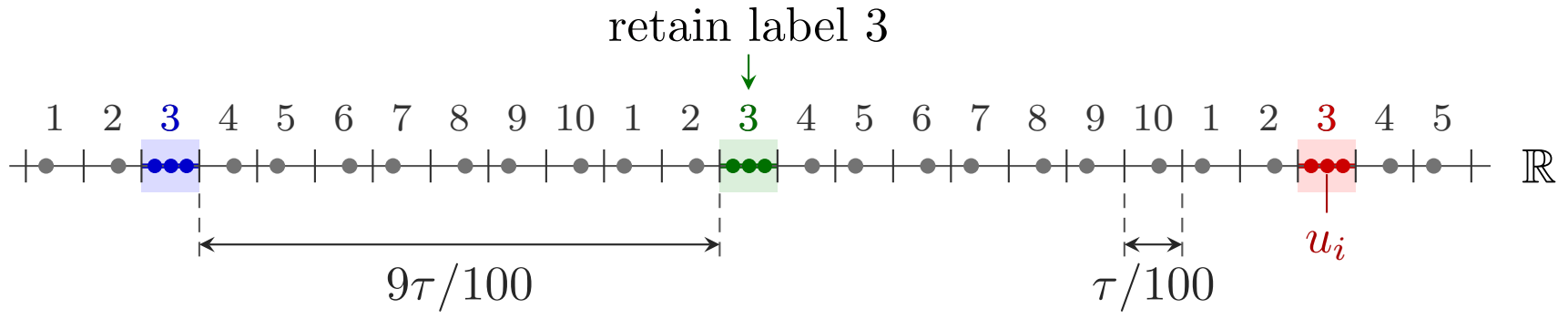

We then retain only the objects whose $u_i$-values lie in intervals of the label that contains the most $u_i$-values. Objects whose $u_i$-values lie in the same interval form a \emph{slope group}. This retains at least a tenth of the objects, so it suffices to bound the number of retained objects: any such bound also bounds $|\mathcal P|$, up to a factor of $10$. Moreover, two intervals with the same label are always separated by a gap of at least $9\tau/100$.

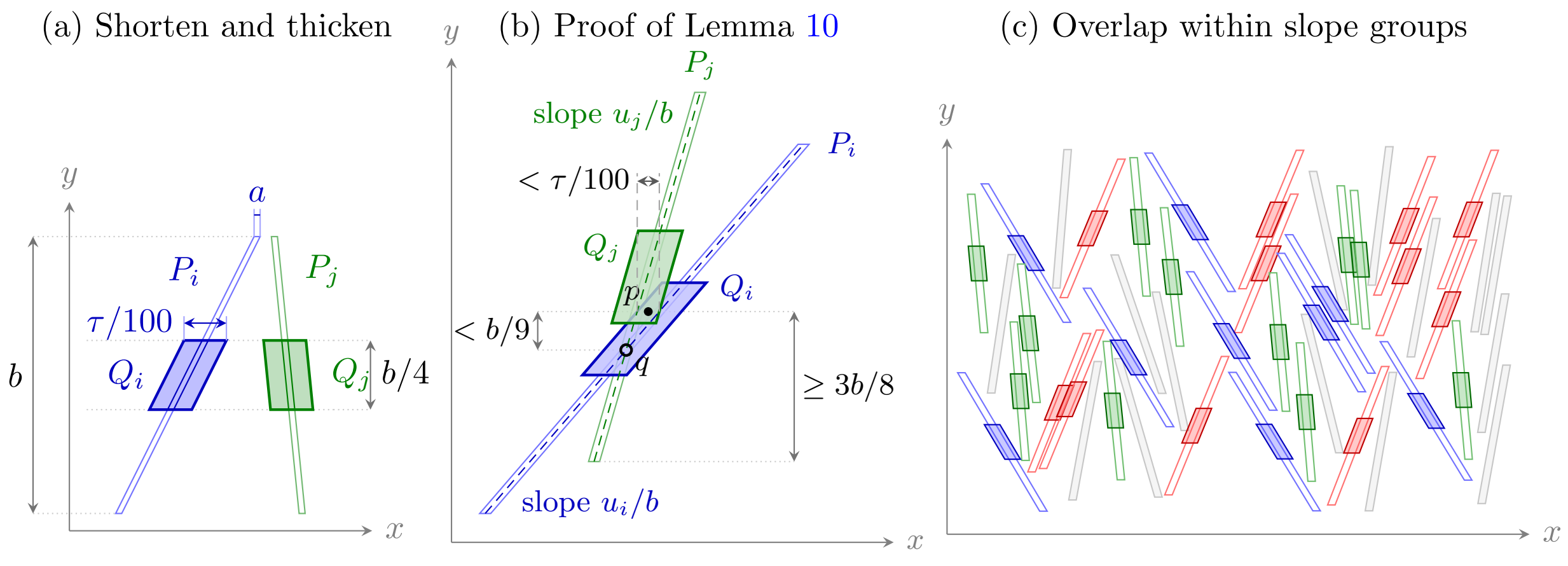
\begin{figure}[h]
    \centering
    \begin{tikzpicture}[>=stealth,every node/.style={font=\small}]
        \begin{scope}[scale=.9]
            \draw[->,black!50] (-.2,-.2)--(3.3,-.2) node[right] {$x$};
            \draw[->,black!50] (-.2,-.2)--(-.2,3.6) node[above] {$y$};
            \draw[blue!55,thin] (.335,0)--(.405,0)--(2.005,3.2)--(1.935,3.2)--cycle;
            \draw[green!50!black!55,thin] (2.455,0)--(2.525,0)--(2.205,3.2)--(2.135,3.2)--cycle;
            \draw[black!20,densely dotted] (0,1.2)--(2.85,1.2) (0,2)--(2.85,2);
            \filldraw[blue!75!black,fill=blue!25,thick]
                (.725,1.2)--(1.215,1.2)--(1.615,2)--(1.125,2)--cycle;
            \filldraw[green!50!black,fill=green!50!black!25,thick]
                (2.125,1.2)--(2.615,1.2)--(2.535,2)--(2.045,2)--cycle;
            \draw[blue!75!black,thin] (.935,1.2)--(1.335,2) (1.005,1.2)--(1.405,2);
            \draw[green!50!black,thin] (2.335,1.2)--(2.255,2) (2.405,1.2)--(2.325,2);
            \node[text=blue!75!black,anchor=east] at (1.47,2.8) {$P_i$};
            \node[text=green!50!black,anchor=west] at (2.37,2.8) {$P_j$};
            \node[text=blue!75!black,anchor=east] at (.83,1.6) {$Q_i$};
            \node[text=green!50!black,anchor=west] at (2.68,1.6) {$Q_j$};
            \draw[blue!35,thin] (1.935,3.2)--(1.935,3.52) (2.005,3.2)--(2.005,3.52);
            \draw[blue!75!black,thin] (1.935,3.45)--(2.005,3.45);
            \node[above,inner sep=2pt,text=blue!75!black] at (1.97,3.52) {$a$};
            \draw[blue!35,thin] (1.125,2)--(1.125,2.28) (1.615,2)--(1.615,2.28);
            \draw[<->,blue!75!black,thin] (1.125,2.2)--(1.615,2.2);
            \node[anchor=east,inner sep=2pt,text=blue!75!black] at (1.08,2.2) {$\tau/100$};
            \draw[black!20,densely dotted] (-.62,0)--(.335,0) (-.62,3.2)--(1.935,3.2);
            \draw[<->,black!55,thin] (-.62,0)--(-.62,3.2);
            \node[left,inner sep=3pt] at (-.62,1.6) {$b$};
            \draw[black!20,densely dotted] (2.85,1.2)--(3.28,1.2) (2.85,2)--(3.28,2);
            \draw[<->,black!55,thin] (3.28,1.2)--(3.28,2);
            \node[right,inner sep=3pt] at (3.28,1.6) {$b/4$};
        \end{scope}
        \begin{scope}[xshift=3.45cm,x=2.3cm,y=1.2cm,every node/.style={font=\footnotesize}]
            \draw[->,black!50] (.15,-.25)--(2.15,-.25) node[right] {$x$};
            \draw[->,black!50] (.15,-.25)--(.15,3.95) node[above] {$y$};
            \draw[blue!55,thin] (.2775,0)--(.3275,0)--(1.7675,3.2)--(1.7175,3.2)--cycle;
            \draw[green!50!black!55,thin] (.77,.45)--(.82,.45)--(1.30,3.65)--(1.25,3.65)--cycle;
            \draw[black!20,densely dotted] (.6,1.75)--(1.7,1.75) (.6,1.4167)--(.89,1.4167) (.82,.45)--(1.7,.45);
            \filldraw[blue!75!black,fill=blue!25,fill opacity=.8,thick]
                (.7425,1.2)--(.9425,1.2)--(1.3025,2)--(1.1025,2)--cycle;
            \filldraw[green!50!black,fill=green!50!black!25,fill opacity=.8,thick]
                (.875,1.65)--(1.075,1.65)--(1.195,2.45)--(.995,2.45)--cycle;
            \draw[blue!75!black,densely dashed,thin] (.3025,0)--(1.7425,3.2);
            \draw[green!50!black,densely dashed,thin] (.795,.45)--(1.275,3.65);
            \fill (1.04,1.75) circle (1.3pt);
            \node[above left,inner sep=1pt] at (1.02,1.77) {$p$};
            \draw[thick] (.94,1.4167) circle (1.6pt);
            \node[below right,inner sep=1pt] at (.96,1.39) {$q$};
            \draw[black!35,thin,densely dashed] (.99,1.75)--(.99,2.95) (1.09,1.75)--(1.09,2.95);
            \draw[<->,black!65,thin] (.99,2.88)--(1.09,2.88);
            \node[left,inner sep=2pt] at (.99,2.88) {$<\tau/100$};
            \draw[black!20,densely dotted] (.54,1.4167)--(.6,1.4167) (.54,1.75)--(.6,1.75);
            \draw[<->,black!55,thin] (.54,1.4167)--(.54,1.75);
            \node[left,inner sep=2pt] at (.54,1.58) {$<b/9$};
            \draw[<->,black!55,thin] (1.7,.45)--(1.7,1.75);
            \node[right,inner sep=2pt] at (1.7,1.1) {$\geq3b/8$};
            \node[text=blue!75!black,anchor=west] at (1.79,3.2) {$P_i$};
            \node[text=green!50!black,above] at (1.275,3.65) {$P_j$};
            \node[text=blue!75!black,anchor=north west,inner sep=1pt] at (.45,.25) {slope $u_i/b$};
            \node[text=green!50!black,anchor=east] at (1.21,3.45) {slope $u_j/b$};
            \node[text=blue!75!black,anchor=west,inner sep=2pt] at (1.33,1.95) {$Q_i$};
            \node[text=green!50!black,anchor=east,inner sep=2pt] at (.95,2.3) {$Q_j$};
        \end{scope}
        \begin{scope}[xshift=8.95cm,scale=1.2]
            \draw[->,black!50] (0,-.18)--(5.05,-.18) node[right] {$x$};
            \draw[->,black!50] (0,-.18)--(0,3.25) node[above] {$y$};
            \foreach \xx/\yy/\u in {
                3.988/1.357/0.222,
                1.833/1.041/-0.412,
                1.055/0.117/0.220,
                0.350/1.044/0.179,
                4.541/1.575/0.189,
                4.600/0.703/0.206,
                4.627/1.552/0.200,
                4.602/0.086/0.216,
                3.369/0.526/-0.298,
                1.800/0.142/-0.348,
                2.029/0.719/-0.251,
                3.405/1.765/-0.265,
                0.915/1.954/0.098,
                3.625/0.023/0.208,
                3.649/1.979/0.153} {
                \filldraw[draw=black!22,fill=black!4,thin]
                    (\xx,\yy)--({\xx+.065},\yy)--({\xx+.065+\u},{\yy+1.2})--({\xx+\u},{\yy+1.2})--cycle;
            }
            \foreach \col/\xx/\yy/\u in {
                red/0.7/0.36/0.48,
                red/0.8/0.39/0.495,
                green!55!black/3.5/1.64/-0.12,
                green!55!black/3.6/1.6/-0.105,
                blue/3.646/1.216/-0.713,
                green!55!black/1.475/0.292/-0.122,
                red/3.718/1.543/0.467,
                blue/1.9/0.752/-0.718,
                green!55!black/0.648/0.46/-0.102,
                red/2.489/1.568/0.489,
                blue/2.434/1.924/-0.719,
                green!55!black/1.988/1.45/-0.139,
                red/0.98/1.871/0.499,
                blue/1.007/1.657/-0.711,
                green!55!black/0.289/1.567/-0.112,
                red/1.887/0.161/0.491,
                blue/3.171/0.023/-0.729,
                green!55!black/3.128/0.279/-0.101,
                red/3.299/0.074/0.467,
                blue/4.485/0.289/-0.73,
                green!55!black/4.489/0.47/-0.11,
                red/4.065/1.17/0.461,
                blue/2.795/0.895/-0.72,
                green!55!black/1.698/1.884/-0.113,
                red/2.529/1.948/0.498,
                blue/0.803/0.02/-0.711,
                green!55!black/2.906/0.886/-0.113,
                red/4.252/1.948/0.463,
                blue/3.563/1.055/-0.701,
                green!55!black/0.743/0.962/-0.133,
                red/3.699/1.893/0.469} {
                \draw[\col!55,thin]
                    (\xx,\yy)--({\xx+.065},\yy)--({\xx+.065+\u},{\yy+1.2})--({\xx+\u},{\yy+1.2})--cycle;
                \filldraw[draw=\col!75!black,fill=\col!70,fill opacity=.3,thin]
                    ({\xx+.375*\u-.035},{\yy+.45})--({\xx+.375*\u+.100},{\yy+.45})--({\xx+.625*\u+.100},{\yy+.75})--({\xx+.625*\u-.035},{\yy+.75})--cycle;
            }
        \end{scope}
        \node at (1.35,5.05) {(a) Shorten and thicken};
        \node at (6.05,5.05) {(b) Proof of \cref{lem:slope-disjoint}};
        \node at (11.93,5.05) {(c) Overlap within slope groups};
    \end{tikzpicture}
    \caption{Intersections of the objects with a plane parallel to the $xy$-plane. (a) Original objects $P_i,P_j$ and their thickened middle portions $Q_i,Q_j$. (b) If $Q_i$ and $Q_j$ share a point $p$, the center lines of $P_i$ and $P_j$ (dashed) are less than $\tau/100$ apart at $p$, so they cross at a point $q$ less than $b/9$ from $p$ in $y$, which lies inside both parallelograms, leading to a contradiction. (c) Grey objects are discarded, and overlaps occur only within a slope group.}
    \label{fig:parameter-groups}
\end{figure}

For every retained object, define the \emph{auxiliary body}

\begin{equation}
\label{eq:auxiliary-body}
Q_i=p_i+\left\{
\lambda_1\begin{pmatrix}\tau/100\\0\\0\end{pmatrix}
+\lambda_2\frac{1}{4}\begin{pmatrix}u_i\\b\\0\end{pmatrix}
+\lambda_3\begin{pmatrix}v_i\\k_i b\\c\end{pmatrix}
:\lambda_1,\lambda_2,\lambda_3\in[-\tfrac12,\tfrac12]
\right\}.
\end{equation}
Thus $Q_i$ has the same center as $P_i$, a first edge of length $\tau/100$, and one quarter of its second edge. In each plane parallel to the $xy$-plane, this is precisely the shortening and thickening shown in \cref{fig:parameter-groups}. Its volume is $bc\tau/400$.

\begin{lemma}
\label{lem:slope-disjoint}
Auxiliary bodies from different retained slope groups have disjoint interiors. Every $Q_i$ lies in $[-\tau/200,1+\tau/200]\times[0,1]^2$.
\end{lemma}

\begin{proof}

Suppose, for a contradiction, that $Q_i$ and $Q_j$ from different retained groups share an interior point $p$. In the plane through $p$ parallel to the $xy$-plane, each auxiliary body is a thickened middle quarter of the corresponding parallelogram section of $P_i$ or $P_j$; see \cref{fig:parameter-groups}(a). The center lines of the original sections have slopes $u_i/b$ and $u_j/b$. Since $p$ lies in both thickened sections, the lines are less than $\tau/100$ apart in the $x$-direction at $p$'s $y$-coordinate. Their slopes differ by $|u_i-u_j|/b\geq9\tau/(100b)$, so they intersect at a point $q$ within a $y$-distance of $b/9$ from $p$; see \cref{fig:parameter-groups}(b).

By \eqref{eq:auxiliary-body}, shortening retains the middle quarter of each original section in the $y$-direction, so $p$ is at least $3b/8$ from either end in that direction. Since $b/9<3b/8$, the intersection point $q$ lies inside both original sections and is therefore an interior point of both $P_i$ and $P_j$. Their interiors thus overlap, a contradiction to the assumption that the original objects form a packing with pairwise disjoint interiors. Finally, shortening preserves containment in the unit cube, and thickening adds less than $\tau/200$ on either side in the $x$-direction, proving the stated containment.

\end{proof}

\paragraph{Finding Tubes}
For a retained interval $I$ with midpoint $u_I$, define
\[
T_I(x,y,z)=(bx-u_Iy,\tau y,z/c)\]

which has 
$|\det T_I|=b\tau/c.$
Applying this map to \eqref{eq:auxiliary-body}, gives
\begin{equation}
\label{eq:transformed-auxiliary-body}
T_I(Q_i)=T_I(p_i)+\left\{
\lambda_1\begin{pmatrix}b\tau/100\\0\\0\end{pmatrix}
+\frac{\lambda_2}{4}\begin{pmatrix}b(u_i-u_I)\\b\tau\\0\end{pmatrix}
+\lambda_3\begin{pmatrix}b(v_i-k_i u_I)\\k_i b\tau\\1\end{pmatrix}
:\lambda_1,\lambda_2,\lambda_3\in[-\tfrac12,\tfrac12]
\right\}.
\end{equation}

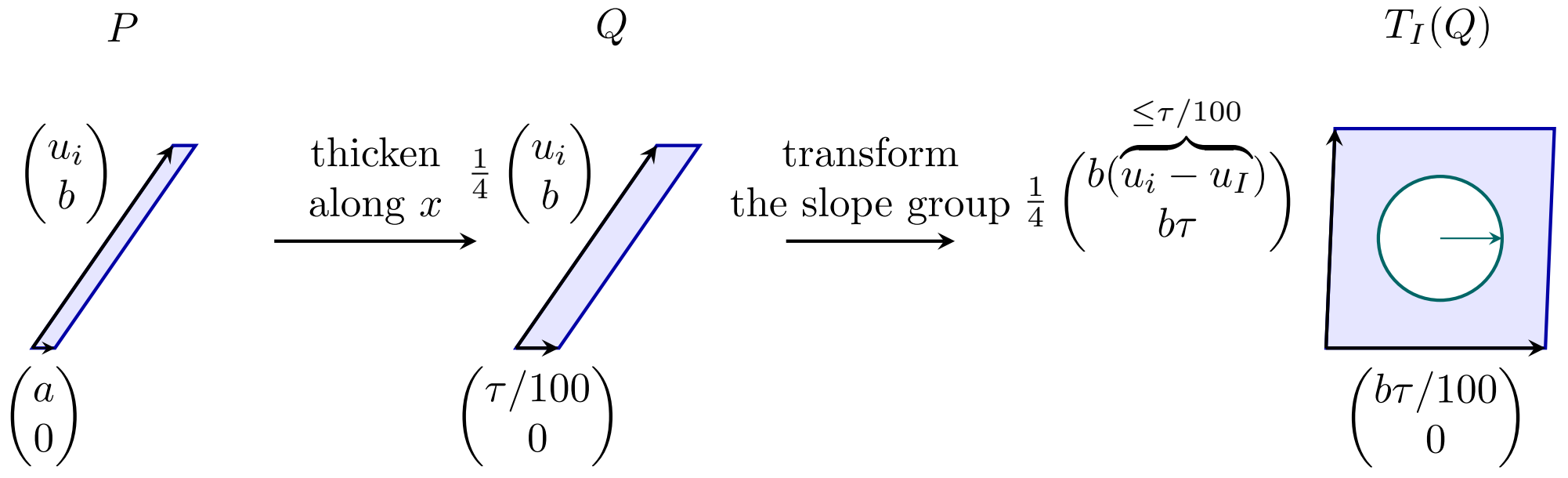
\begin{figure}[h]
    \centering
    \begin{tikzpicture}[
        scale=0.95,
        >=stealth,
        body/.style={draw=blue!65!black,fill=blue!10,thick},
        every node/.style={font=\small}
    ]
        \begin{scope}
            \node at (0.8,2.85) {$P$};
            \draw[body]
                (0,0)--(0.20,0)--(1.45,1.8)--(1.25,1.8)--cycle;
            \draw[->,black,thin] (0,0)--(0.20,0);
            \draw[->,black,thick] (0,0)--(1.25,1.8);
            \node[below] at (0.10,-0.02) {$\begin{pmatrix}a\\0\end{pmatrix}$};
            \node[anchor=east,inner sep=2pt] at (.82,1.55) {$\begin{pmatrix}u_i\\b\end{pmatrix}$};
        \end{scope}
        \draw[->,thick] (2.15,0.95)--(3.95,0.95)
            node[midway,above,align=center] {thicken\\along $x$};
        \begin{scope}[xshift=4.3cm]
            \node at (0.85,2.85) {$Q$};
            \draw[body]
                (0,0)--(0.38,0)--(1.63,1.8)--(1.25,1.8)--cycle;
            \draw[->,black,thick] (0,0)--(0.38,0);
            \draw[->,black,thick] (0,0)--(1.25,1.8);
            \node[below] at (0.19,-0.02) {$\begin{pmatrix}\tau/100\\0\end{pmatrix}$};
            \node[anchor=east,inner sep=2pt] at (.82,1.55) {$\frac14\begin{pmatrix}u_i\\b\end{pmatrix}$};
        \end{scope}
        \draw[->,thick] (6.7,0.95)--(8.2,0.95)
            node[midway,above,align=center] {transform\\the slope group};
        \begin{scope}[xshift=11.5cm]
            \node at (1,2.85) {$T_I(Q)$};
            \draw[body]
                (0,0)--(1.95,0)--(2.03,1.95)--(0.08,1.95)--cycle;
            \draw[teal!80!black,thick,fill=white]
                (1.015,0.975) circle[radius=0.55];
            \draw[->,teal!80!black] (1.015,0.975)--(1.565,0.975);
            \node[below] at (.975,-.03)
                {$\begin{pmatrix}b\tau/100\\0\end{pmatrix}$};
            \node[anchor=east,inner sep=2pt] at (-.16,1.3)
                {$\frac14\begin{pmatrix}b(\smash{\overbrace{u_i-u_I}^{\le\tau/100}})\\b\tau\end{pmatrix}$};
            \draw[->,black,thick] (0,0)--(1.95,0);
            \draw[->,black,thick] (0,0)--(0.08,1.95);
        \end{scope}
    \end{tikzpicture}
    \caption{The disk in $T_I(Q)$ has radius $\Theta(b\tau)$ and yields
    a tube along the object. Arrows show the generating edges, labeled
    with their $xy$-components;
    the drawings are schematic and not to scale.}
    \label{fig:tube-transformation}
\end{figure}

Let $D_i$ denote the third edge vector in \eqref{eq:transformed-auxiliary-body}, with unit direction $d_i=D_i/\|D_i\|$, and set $\delta=b\tau/1000$. The following lemma shows that each $2T_I(Q_i)$ contains a tube of radius $\delta$ in direction $d_i$, with the tube directions pairwise $\delta$-separated within each retained slope group.

\begin{lemma}
\label{lem:separated-tubes}
Within each retained slope group, every $2T_I(Q_i)$ contains the tube $S_i=T_\delta(d_i,2T_I(p_i))$, and the unoriented angle between $d_i$ and $d_j$ is at least $\delta$ for each $i\neq j$.
\end{lemma}

\begin{proof}
The first two edge vectors in \eqref{eq:transformed-auxiliary-body} span a parallelogram in the $xy$-plane of $x$-width $b\tau/100$ and $y$-height $b\tau/4$. Since $u_i\in I$, the shear satisfies
\[
\left|\frac{b(u_i-u_I)}4\right|\leq\frac{b\tau}{800}.
\]
This bound on the shear implies that the parallelogram contains the disk $B$ centered at the origin with radius $b\tau/400$; see \cref{fig:tube-transformation}. Moreover, $1\leq\|D_i\|\leq4$ follows from $|v_i|\leq1$, $|k_i|b\leq1$, $|u_I|\leq1.01$, and $\tau\leq2$. After translating so that $T_I(p_i)=0$, \eqref{eq:transformed-auxiliary-body} gives
\[
T_I(Q_i)\supseteq\bigl\{q+\lambda D_i:q\in B,\ |\lambda|\leq\tfrac12\bigr\}.
\]
To prove the tube containment, it suffices to show that $T_I(Q_i)$ contains $\frac12T_\delta(d_i,0)$, the $\delta/2$-neighbourhood of the segment $\{td_i:|t|\leq\frac14\}$. Any point in this set can be written as $td_i+e$, where $|t|\leq\frac14$ and $\|e\|<\delta/2$. Writing $e_z$ for the $z$-coordinate of $e$, we have
\[
td_i+e=(e-e_zD_i)+\Bigl(\frac{t}{\|D_i\|}+e_z\Bigr)D_i.
\]
Because $D_i$ has $z$-coordinate $1$, the vector $e-e_zD_i$ lies in the $xy$-plane and satisfies
\[
\|e-e_zD_i\|\leq\|e\|+|e_z|\,\|D_i\|
<\frac{5\delta}{2}=\frac{b\tau}{400}.
\]
Thus $e-e_zD_i\in B$, while the coefficient of $D_i$ has absolute value at most $\frac14+\frac\delta2\leq\frac12$. The point therefore lies in $T_I(Q_i)$, proving the claimed containment.

For the angular separation, take distinct objects $i,j$ in the same group. If $k_i\neq k_j$, the second coordinates of $D_i$ and $D_j$ differ by at least $b\tau$. If $k_i=k_j$, the separation assumption and $|u_i-u_j|<\tau/100$ give $|v_i-v_j|>99\tau/100$, so their first coordinates differ by more than $99b\tau/100$. In either case,
\[
\|D_i-D_j\|\geq\frac{99b\tau}{100}.
\]
Let $\varphi$ be the unoriented angle between $d_i$ and $d_j$, and consider the triangle with vertices $0$, $D_i$, and $D_j$. Its area is $\frac12\|D_i\|\,\|D_j\|\sin\varphi\leq8\sin\varphi$. The line through $D_i$ and $D_j$ lies in the plane $z=1$, so its distance from the origin is at least $1$ and the area is at least $\frac12\|D_i-D_j\|$. Comparing these bounds gives
\[
\varphi\geq\sin\varphi\geq\frac{\|D_i-D_j\|}{16}
\geq\frac{99b\tau}{1600}\geq\delta.
\]

\end{proof}

\subsection{The volume bound}

Fix a retained interval $I$ containing $n_I$ objects. By \cref{lem:separated-tubes}, the tubes $S_i$ with $u_i\in I$ have pairwise $\delta$-separated directions, so \cref{thm:tube-union} applies with $\delta=b\tau/1000\leq1$. Using $S_i\subseteq2T_I(Q_i)$, $\operatorname{vol}(S_i)\geq\pi\delta^2$, and $|\det(2T_I)|=8b\tau/c$, we obtain

\[
\begin{aligned}
\operatorname{vol}\!\left(\bigcup_{u_i\in I}Q_i\right)
&=\frac{c}{8b\tau}\,
  \operatorname{vol}\!\left(\bigcup_{u_i\in I}2T_I(Q_i)\right)
\geq\frac{c}{8b\tau}\,
  \operatorname{vol}\!\left(\bigcup_{u_i\in I}S_i\right)\\
&\geq\frac{c}{8b\tau}\cdot\frac{\delta^\varepsilon}{C_\varepsilon}
  \sum_{u_i\in I}\operatorname{vol}(S_i)
\geq\frac{c}{8b\tau}\cdot\frac{\delta^\varepsilon}{C_\varepsilon}\,
  n_I\pi\delta^2
\geq\frac{n_I\,bc\tau\,\delta^\varepsilon}{3\cdot10^6\,C_\varepsilon}.
\end{aligned}
\]
By \cref{lem:slope-disjoint}, the unions for distinct retained intervals have disjoint interiors and lie in a common box of volume $1+\tau/100\leq2$. Summing the preceding bound over the retained intervals therefore gives
\[
2\geq\operatorname{vol}\Bigl(\bigcup_{\text{retained }i}Q_i\Bigr)=\sum_I\operatorname{vol}\Bigl(\bigcup_{u_i\in I}Q_i\Bigr)\geq\frac{|\mathcal P'|\,bc\tau\,\delta^\varepsilon}{3\cdot10^7\,C_\varepsilon},
\]
where the sum runs over the retained intervals, which together contain at least $|\mathcal P'|/10$ objects. Rearranging this inequality gives
\[
|\mathcal P'|\leq6\cdot10^7\,C_\varepsilon\,\frac{\delta^{-\varepsilon}}{bc\tau}\leq6\cdot10^{10}\,C_\varepsilon\,\frac{n^{2\varepsilon}}{bc\tau},
\]
where the last step assumes $b\tau\geq n^{-2}$ and $\varepsilon\leq1$, so that $\delta^{-\varepsilon}\leq1000\,(b\tau)^{-\varepsilon}\leq1000\,n^{2\varepsilon}$.

Otherwise, \cref{thm:counting} holds trivially, since $|\mathcal P|\leq n<1/(b\tau)\leq1/(bc\tau)$.
Adding the at most $100/(bc\tau)$ objects with $a_i\geq\tau/100$ to the bound on $\mathcal P'$, replacing $\varepsilon$ by $\varepsilon/2$, and absorbing the constants into $C_\varepsilon$ proves \cref{thm:counting}.

\section{Reduction to \texorpdfstring{$bc$}{bc}-parallelepipeds}
\label{sec:reduction}

In this section, we reduce general convex polyhedra $K_1,\ldots,K_n$ to the $bc$-parallelepipeds of \cref{sec:algorithm}. \Cref{sec:enclose} constructs enclosures only a constant factor larger than the objects, with polylogarithmically many pairs $(b,c)$ defining the groups packed by \cref{alg:canonical-packing}. \Cref{sec:proof-main} gives the complete algorithm: we guess the dimensions of an optimal container, pack each group with \cref{alg:canonical-packing}, handle very thin objects separately, and combine the packings into one box. Throughout, $w_r(K)$ denotes the width of a polyhedron $K$ in coordinate $r\in\{x,y,z\}$, and in \cref{sec:enclose} we assume that all these widths are at most $1$.

\subsection{Enclosing each object}\label{sec:enclose}

\begin{figure}[h]
    \centering
    \begin{tikzpicture}[>=stealth,every node/.style={font=\small},
        first/.style={red!75!black,very thick,->},
        second/.style={blue!80!black,very thick,->},
        third/.style={green!50!black,very thick,->}]
        \draw[black!50,densely dashed,cm={1.0000,0.0000,0.0000,1.0000,(0.000,0)}] (0,0) circle (1.7321);
        \draw[blue!55!black,densely dashed,thin] (-1.412,-0.965)--(-0.074,-0.457);
        \draw[blue!55!black,densely dashed,thin] (-0.074,-0.457)--(-0.074,1.423);
        \draw[blue!55!black,densely dashed,thin] (-0.074,-0.457)--(1.412,-0.914);
        \shade[ball color=black!35,opacity=.85,cm={1.0000,0.0000,0.0000,1.0000,(0.000,0)}] (0,0) circle (1);
        \draw[blue!70!black,thick] (-1.412,0.914)--(-0.074,1.423);
        \draw[blue!70!black,thick] (-1.412,-0.965)--(-1.412,0.914);
        \draw[blue!70!black,thick] (0.074,-1.423)--(1.412,-0.914);
        \draw[blue!70!black,thick] (1.412,-0.914)--(1.412,0.965);
        \draw[blue!70!black,thick] (0.074,0.457)--(1.412,0.965);
        \draw[blue!70!black,thick] (0.074,-1.423)--(0.074,0.457);
        \draw[blue!70!black,thick] (-1.412,-0.965)--(0.074,-1.423);
        \draw[blue!70!black,thick] (-1.412,0.914)--(0.074,0.457);
        \draw[blue!70!black,thick] (-0.074,1.423)--(1.412,0.965);
        \node[text=black] at (0.000,0.000) {$B$};
        \node[text=blue!70!black,anchor=south west,inner sep=1pt] at (1.412,0.965) {$C$};
        \node[text=black!60,anchor=north east,inner sep=1pt] at (-1.327,-1.113) {$\sqrt3B$};
        \draw[->,black!60,thick] (1.962,0)--(3.012,0) node[midway,above] {$A$};
        \draw[black!50,densely dashed,cm={1.1991,0.2927,0.0000,0.9285,(6.600,0)}] (0,0) circle (1.7321);
        \draw[blue!55!black,densely dashed,thin] (6.192,-0.537)--(7.071,1.325);
        \draw[blue!55!black,densely dashed,thin] (6.192,-0.537)--(7.738,-1.013);
        \draw[second,densely dashed] (4.583,-0.849)--(6.192,-0.537);
        \shade[ball color=black!35,opacity=.85,cm={1.1991,0.2927,0.0000,0.9285,(6.600,0)}] (0,0) circle (1);
        \draw[blue!70!black,thick] (5.462,1.013)--(7.071,1.325);
        \draw[blue!70!black,thick] (6.129,-1.325)--(7.738,-1.013);
        \draw[blue!70!black,thick] (7.738,-1.013)--(8.617,0.849);
        \draw[blue!70!black,thick] (7.008,0.537)--(8.617,0.849);
        \draw[blue!70!black,thick] (6.129,-1.325)--(7.008,0.537);
        \draw[blue!70!black,thick] (5.462,1.013)--(7.008,0.537);
        \draw[blue!70!black,thick] (7.071,1.325)--(8.617,0.849);
        \draw[first] (4.583,-0.849)--(6.129,-1.325) node[pos=.5,below,font=\footnotesize] {$(a,0,0)$};
        \node[above,font=\footnotesize,text=blue!80!black,inner sep=2pt] at (5.387,-0.693) {$(u,b,0)$};
        \draw[third] (4.583,-0.849)--(5.462,1.013) node[pos=.5,left,font=\footnotesize] {$(v,q,c)$};
        \node[text=black] at (6.600,0.000) {$E$};
        \node[text=blue!70!black,anchor=south west,inner sep=1pt] at (8.617,0.849) {$F$};
        \node[text=black!60,anchor=north west,inner sep=1pt] at (8.399,-0.365) {$\sqrt3E$};
        \node at (0.000,-2.132) {(a) Ball and cube};
        \node at (6.600,-2.132) {(b) Ellipsoid and parallelepiped};
        \begin{scope}[xshift=0.000cm,yshift=-4.431cm]
            \fill[blue!6] (-2.017,-0.849)--(-0.408,-0.537)--(0.471,1.325)--(-1.138,1.013)--cycle;
            \fill[blue!6] (-0.408,-0.537)--(1.138,-1.013)--(2.017,0.849)--(0.471,1.325)--cycle;
            \fill[blue!6] (-2.017,-0.849)--(-0.471,-1.325)--(1.138,-1.013)--(-0.408,-0.537)--cycle;
            \draw[black!35,densely dashed,thin] (-0.408,-0.537)--(0.471,1.325);
            \draw[black!35,densely dashed,thin] (-0.408,-0.537)--(1.138,-1.013);
            \draw[second,densely dashed] (-2.017,-0.849)--(-0.408,-0.537);
            \filldraw[fill=black!16,draw=black!65,line join=round] (0.285,0.809)--(1.196,0.641)--(0.808,0.828)--(-0.102,0.995)--cycle;
            \filldraw[fill=black!32,draw=black!65,line join=round] (-0.932,0.182)--(-0.720,-0.729)--(0.346,-0.249)--(0.134,0.662)--cycle;
            \filldraw[fill=black!42,draw=black!65,line join=round] (0.480,-0.274)--(0.281,-0.942)--(1.039,-0.207)--(1.238,0.462)--cycle;
            \filldraw[fill=black!32,draw=black!65,line join=round] (-0.932,0.182)--(-0.997,0.213)--(-1.220,-0.537)--(-1.204,-0.608)--(-0.749,-0.826)--(-0.720,-0.729)--cycle;
            \filldraw[fill=black!42,draw=black!65,line join=round] (-0.720,-0.729)--(-0.749,-0.826)--(0.166,-0.994)--(0.281,-0.942)--(0.480,-0.274)--(0.346,-0.249)--cycle;
            \filldraw[fill=black!11,draw=black!65,line join=round] (-0.102,0.995)--(-0.269,0.920)--(-0.997,0.213)--(-0.932,0.182)--(0.134,0.662)--(0.285,0.809)--cycle;
            \filldraw[fill=black!37,draw=black!65,line join=round] (0.285,0.809)--(0.134,0.662)--(0.346,-0.249)--(0.480,-0.274)--(1.238,0.462)--(1.196,0.641)--cycle;
            \draw[black!65,thin] (-1.138,1.013)--(0.471,1.325);
            \draw[black!65,thin] (-0.471,-1.325)--(1.138,-1.013);
            \draw[black!65,thin] (1.138,-1.013)--(2.017,0.849);
            \draw[black!65,thin] (0.408,0.537)--(2.017,0.849);
            \draw[black!65,thin] (-0.471,-1.325)--(0.408,0.537);
            \draw[black!65,thin] (-1.138,1.013)--(0.408,0.537);
            \draw[black!65,thin] (0.471,1.325)--(2.017,0.849);
            \draw[first] (-2.017,-0.849)--(-0.471,-1.325) node[pos=.5,below,font=\footnotesize] {$(a,0,0)$};
            \node[below left,font=\footnotesize,text=blue!80!black,inner sep=2pt] at (-2.017,-0.849) {$(u,b,0)$};
            \draw[third] (-2.017,-0.849)--(-1.138,1.013) node[pos=.5,left,font=\footnotesize] {$(v,q,c)$};
            \node[text=blue!70!black,above right,inner sep=1pt] at (2.017,0.849) {$F$};
            \node at (0.000,0.147) {$K$};
        \end{scope}
        \begin{scope}[xshift=6.578cm,yshift=-4.431cm]
            \filldraw[fill=black!3,draw=black!55,densely dashed] (-0.991,0.618)--(-1.202,0.413)--(-1.077,-0.933)--(-0.525,-1.147)--(1.021,-0.649)--(1.246,-0.430)--(1.116,0.970)--(0.655,1.149)--cycle;
            \filldraw[fill=blue!6,draw=blue!70!black,thick] (-1.025,-0.675)--(0.575,-0.675)--(1.025,0.675)--(-0.575,0.675)--cycle;
            \filldraw[fill=black!22,draw=black!65] (-0.566,0.353)--(-0.687,0.236)--(-0.615,-0.533)--(-0.300,-0.656)--(0.583,-0.371)--(0.712,-0.246)--(0.638,0.555)--(0.374,0.657)--cycle;
            \fill (0.000,0.000) circle (1.2pt) node[below right,inner sep=1.5pt] {$0$};
            \node[above left,font=\footnotesize,inner sep=1pt] at (0.655,1.149) {$7K$};
            \node[right,text=blue!70!black] at (1.025,0.675) {$F$};
            \node at (-0.220,0.160) {$K$};
        \end{scope}
        \node at (0.000,-6.206) {(c) $K\subseteq F$};
        \node at (6.600,-6.206) {(d) Section through $0$};
    \end{tikzpicture}
    \caption{The enclosure of \cref{lem:flag}. (a,b) Stretching $B\subseteq C\subseteq\sqrt3B$ by $A$ gives $E\subseteq F\subseteq\sqrt3E$. (c) $K$ inside $F$. (d) The section through the center: $K\subseteq F\subseteq7K$ (dashed, not to scale).}
    \label{fig:reduction-enclosure}
\end{figure}
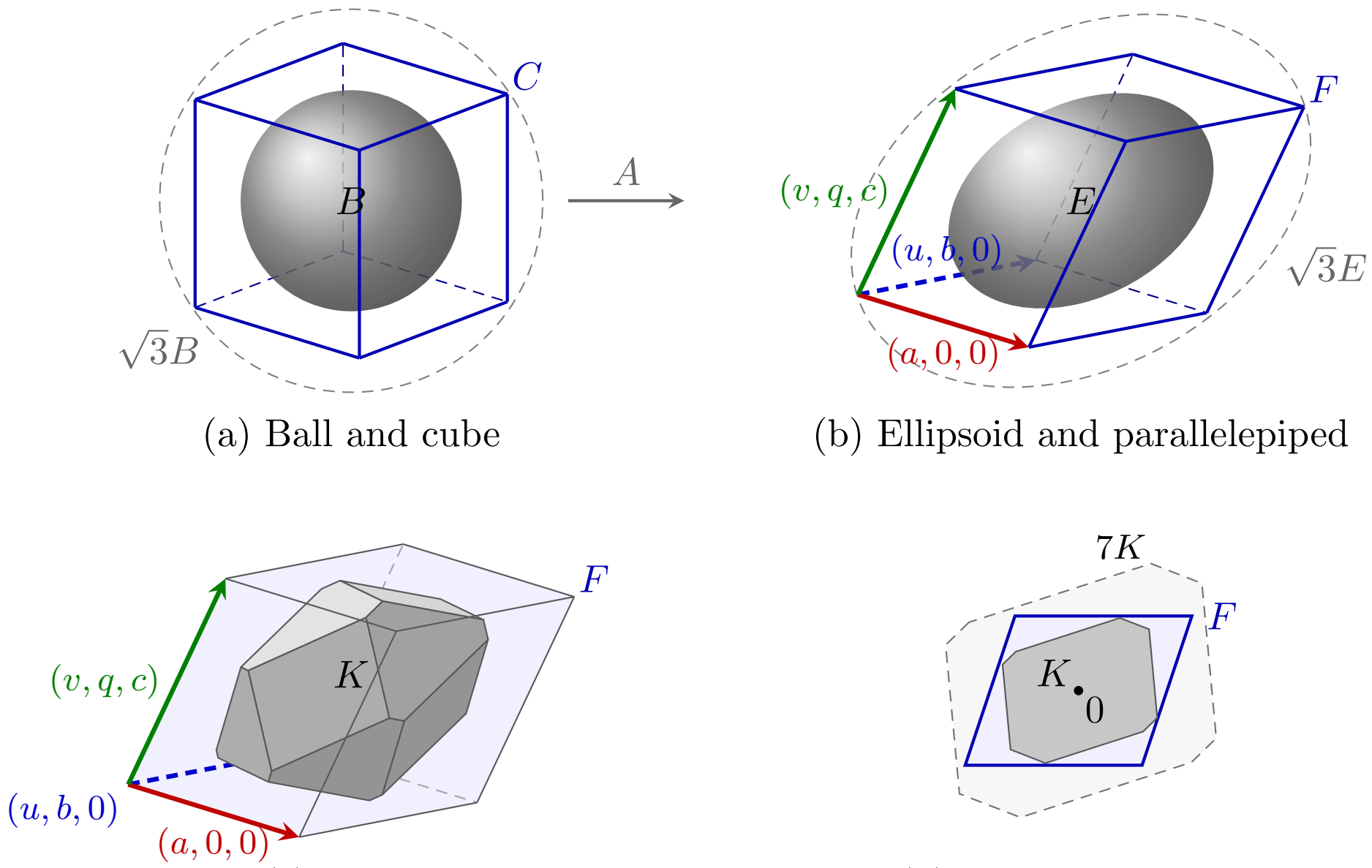

We can enclose each object in a parallelepiped that lies within a constant enlargement of the object; see \cref{fig:reduction-enclosure}.

\begin{lemma}
\label{lem:flag}
Let $K$ be a (full-dimensional) convex polyhedron. In polynomial time, we can find a parallelepiped $F$, with edges $(a,0,0)$, $(u,b,0)$, and $(v,q,c)$, where $a,b,c>0$, such that $K\subseteq F\subseteq7K$.
\end{lemma}

\begin{proof}
By the algorithmic version of John's theorem~\cite{John48} due to Gr\"otschel, Lov\'asz, and Schrijver~\cite{GLS88}, we can, after translating $K$, compute in polynomial time an ellipsoid $E$ centered at the origin such that $\frac14E\subseteq K\subseteq E$.

An ellipsoid is a stretched ball. Stretching a cube around the ball in the same way gives a parallelepiped $F$ around $E$, and since the ball lies in the cube and the cube lies in the ball enlarged by $\sqrt3$, we get $E\subseteq F\subseteq\sqrt3E$; see \cref{fig:reduction-enclosure}(a,b). By rotating the cube before stretching, we can make the edges of $F$ have the form $(a,0,0)$, $(u,b,0)$, and $(v,q,c)$ with $a,b,c>0$. To be precise, let $T$ be the non-singular linear transformation mapping the ball $B$ to the ellipsoid $E$. We first pick a unit vector $e_1$ in the preimage $T^{-1}(\{(x,0,0):x\in \mathbb{R}\})$. Now $T^{-1}(\{(x,y,0):x,y\in \mathbb{R}\})$ has dimension 2, so there exists a vector $e_2$ orthogonal to $e_1$ mapping to this preimage. Finally, picking $e_3$ orthogonal to $e_1$ and $e_2$, the directions $e_1,e_2,e_3$ define the edges of the desired rotated cube.

Since $\frac14E\subseteq K\subseteq E$ and $K$ contains the origin, this gives
\[
K\subseteq E\subseteq F\subseteq\sqrt3E\subseteq4\sqrt3K\subseteq7K.
\]
\end{proof}

For each $K_i$, let $F_i$ be its enclosure from \cref{lem:flag}, with positive edge parameters $a_i,b_i,c_i$ and shears $u_i,v_i,q_i$. If the original objects fit in the unit cube, enlarging that packing by a factor of $7$ allows each $F_i$ to be placed inside the corresponding enlarged $K_i$, so the enclosures fit in a cube of side length $7$ and satisfy
\[
\sum_i a_ib_ic_i=\sum_i\operatorname{vol}(F_i)=O(1).
\]
Their edge coordinates are also $O(1)$, since $F_i\subseteq7K_i$ and all input widths are at most $1$.

We next reduce the number of choices for $b$ and $c$ by rounding $b_i$ and $c_i$ up to the next values in $\{1/n,2/n,4/n,\ldots\}$, denoted $B_i$ and $C_i$, while leaving $a_i$ unchanged. For now, we assume $b_i,c_i\geq1/n$ and handle the remaining objects in the proof of \cref{thm:main}.

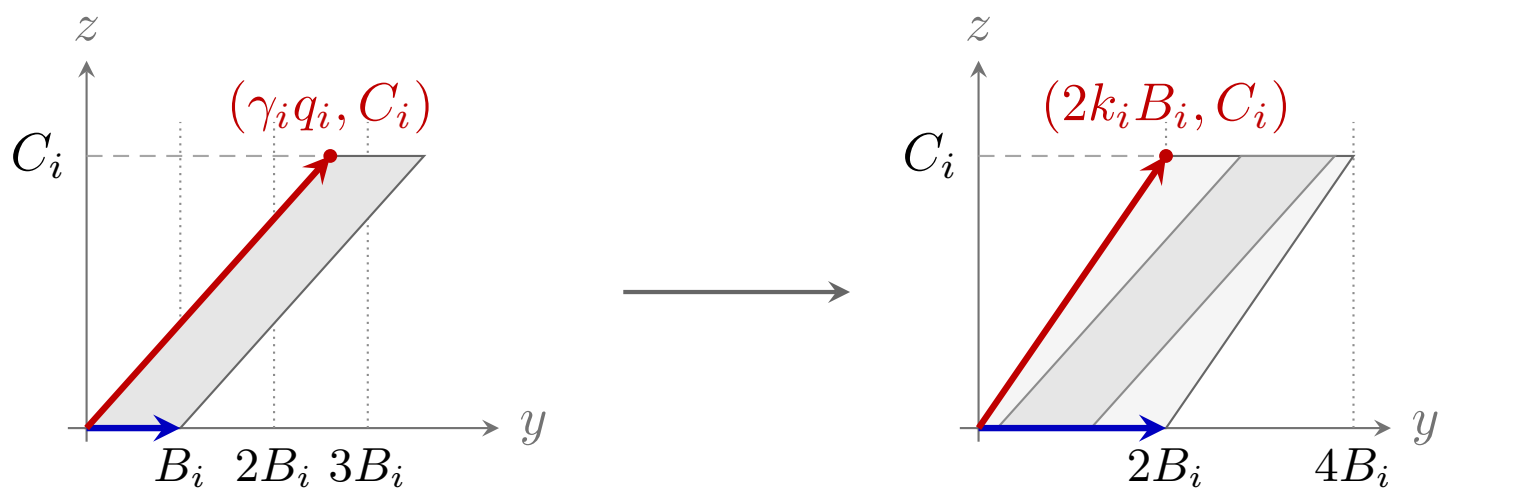
\begin{figure}[h]
    \centering
    \begin{tikzpicture}[>=stealth,every node/.style={font=\small},
        second/.style={blue!75!black,very thick,->},
        third/.style={red!75!black,very thick,->},
        grid/.style={black!45,densely dotted},
        guide/.style={black!35,densely dashed,thin},
        axis/.style={->,black!55,thin}]
        \begin{scope}[x=.62cm,y=.9cm]
            \draw[axis] (-.2,0)--(4.4,0) node[right] {$y$};
            \draw[axis] (0,-.1)--(0,2.7) node[above] {$z$};
            \foreach \j/\lab in {1/B_i,2/2B_i,3/3B_i} {
                \draw[grid] (\j,0)--(\j,2.25);
                \node[below,font=\footnotesize] at (\j,0) {$\lab$};
            }
            \filldraw[fill=black!10,draw=black!60] (0,0)--(1,0)--(3.6,2)--(2.6,2)--cycle;
            \draw[guide] (0,2)--(2.6,2);
            \node[left] at (0,2) {$C_i$};
            \draw[second] (0,0)--(1,0);
            \draw[third] (0,0)--(2.6,2) node[above,text=red!75!black] {$(\gamma_iq_i,C_i)$};
            \fill[red!75!black] (2.6,2) circle (1.3pt);
            \node at (2.1,-.85) {(a) $\gamma_iq_i$ is not a multiple of $B_i$};
        \end{scope}
        \draw[->,black!60,thick] (3.55,.9)--(5.05,.9);
        \begin{scope}[xshift=5.9cm,x=.62cm,y=.9cm]
            \draw[axis] (-.2,0)--(4.4,0) node[right] {$y$};
            \draw[axis] (0,-.1)--(0,2.7) node[above] {$z$};
            \foreach \j/\lab in {2/2B_i,4/4B_i} {
                \draw[grid] (\j,0)--(\j,2.25);
                \node[below,font=\footnotesize] at (\j,0) {$\lab$};
            }
            \filldraw[fill=black!4,draw=black!60] (0,0)--(2,0)--(4,2)--(2,2)--cycle;
            \filldraw[fill=black!10,draw=black!45,thin] (.2,0)--(1.2,0)--(3.8,2)--(2.8,2)--cycle;
            \draw[guide] (0,2)--(2,2);
            \node[left] at (0,2) {$C_i$};
            \draw[second] (0,0)--(2,0);
            \draw[third] (0,0)--(2,2) node[above,text=red!75!black] {$(2k_iB_i,C_i)$};
            \fill[red!75!black] (2,2) circle (1.3pt);
            \node at (2.1,-.85) {(b) $2k_iB_i$ is a multiple of $2B_i$};
        \end{scope}
    \end{tikzpicture}
    \caption{Rounding the $yz$-shadow (gray) of an enclosure. Doubling the base (blue) and moving the slanted side (red) parallel to it makes its $y$-component a multiple of the new base. The new shadow contains the old one.}
    \label{fig:reduction-rounding}
\end{figure}

We now replace $F_i$ by a $bc$-parallelepiped $P_i$ with first edge $f_1$ and $(b,c)=(2B_i,C_i)$. Set $\beta_i=B_i/b_i$ and $\gamma_i=C_i/c_i$, which satisfy $1\leq\beta_i,\gamma_i<2$ because $b_i,c_i\geq1/n$. Scaling $f_2$ and $f_3$ by these factors gives the desired $y$-component $B_i$ and $z$-component $C_i$, but the third edge's $y$-component $\gamma_iq_i$ need not be an integer multiple of $B_i$. We therefore double the second edge and shift the third edge parallel to it; see \cref{fig:reduction-rounding}. Choose an integer $k_i$ nearest to $\gamma_iq_i/(2B_i)$ and set $t_i=\gamma_iq_i/B_i-2k_i$, so that $|t_i|\leq1$. The resulting edges of $P_i$ are
\[
p_1=f_1=\begin{pmatrix}a_i\\0\\0\end{pmatrix},\qquad
p_2=2\beta_if_2=\begin{pmatrix}2\beta_iu_i\\2B_i\\0\end{pmatrix},\qquad
p_3=\gamma_if_3-t_i\beta_if_2=\begin{pmatrix}\gamma_iv_i-t_i\beta_iu_i\\2k_iB_i\\C_i\end{pmatrix}.
\]
The $y$-component of $p_3$ is $k_i$ times that of $p_2$, so $P_i$ is an $bc$-parallelepiped as in \cref{sec:algorithm}, with first edge of length $a_i$ and $(b,c)=(2B_i,C_i)$.

\begin{lemma}
\label{lem:rounded}
We have $K_i\subseteq P_i\subseteq42K_i$.
\end{lemma}

\begin{proof}
By definition,
\[
\begin{aligned}
F_i&=\Bigl\{\mu_1f_1+\mu_2f_2+\mu_3f_3:\mu_1,\mu_2,\mu_3\in[-\tfrac12,\tfrac12]\Bigr\},\\
P_i&=\Bigl\{\lambda_1p_1+\lambda_2p_2+\lambda_3p_3:\lambda_1,\lambda_2,\lambda_3\in[-\tfrac12,\tfrac12]\Bigr\},
\end{aligned}
\]
and scaling $F_i$ by the factor of $6$ gives
\[
6F_i=\Bigl\{\mu_1f_1+\mu_2f_2+\mu_3f_3:\mu_1,\mu_2,\mu_3\in[-3,3]\Bigr\}.
\]
Throughout, we use $1\leq\beta_i,\gamma_i<2$, and $|t_i|\leq1$.

To prove $F_i\subseteq P_i$, we express $f_1,f_2,f_3$ in terms of $p_1,p_2,p_3$:
\[
f_1=p_1,\qquad f_2=\frac{1}{2\beta_i}p_2,\qquad f_3=\frac{1}{\gamma_i}p_3+\frac{t_i}{2\gamma_i}p_2.
\]
Hence a point $\mu_1f_1+\mu_2f_2+\mu_3f_3$ of $F_i$ equals
\[
\mu_1 p_1+\Bigl(\frac{\mu_2}{2\beta_i}+\frac{t_i\mu_3}{2\gamma_i}\Bigr)p_2+\frac{\mu_3}{\gamma_i}\,p_3,
\]
and since $|\mu_1|,|\mu_2|,|\mu_3|\leq\frac12$, these coefficients are at most $\frac12$, $\frac14+\frac14$, and $\frac12$ in absolute value. So the point lies in $P_i$.

To show $P_i\subseteq42K_i$, we first show $P_i\subseteq6F_i$. Note that a point $\lambda_1p_1+\lambda_2p_2+\lambda_3p_3$ of $P_i$, with $|\lambda_1|,|\lambda_2|,|\lambda_3|\leq\frac12$, equals
\[
\lambda_1f_1+\beta_i(2\lambda_2-t_i\lambda_3)\,f_2+\gamma_i\lambda_3\,f_3.
\]
The coefficients of $f_1,f_2,f_3$ have absolute values at most $\frac12$, less than $2\cdot\frac32=3$, and less than $2\cdot\frac12=1$, respectively, so the point lies in $6F_i$. Together with \cref{lem:flag}, this gives $P_i\subseteq6F_i\subseteq42K_i$.
\end{proof}

\FloatBarrier

Objects with the same rounded pair $(B_i,C_i)$ form a type, giving $O(\log^2 n)$ types since both parameters are rounded in powers of two between $1/n$ and a constant. Moreover, $P_i\subseteq42K_i$ by \cref{lem:rounded}, so the $P_i$ have coordinate widths $O(1)$.

\begin{lemma}
\label{lem:type-cost}
Suppose that $K_1,\ldots,K_n$ can be packed into the unit cube. Then, for each type, \cref{alg:canonical-packing} packs the $P_i$ of this type into a box of volume $O_\varepsilon(n^\varepsilon)$.
\end{lemma}

\begin{proof}
Enlarging a packing of the $K_i$ in the unit cube by a factor of $42$ allows us to place each $P_i$ inside the corresponding enlarged $K_i$, since $P_i\subseteq42K_i$ by \cref{lem:rounded}, giving a packing in a cube of side $42$. After scaling by $1/42$, the $P_i$ of each type fit into the unit cube, so \cref{lem:fan,lem:mst}, together with $\sum_ia_ibc\leq1$ and $c\leq1$, give a packing of volume $O_\varepsilon(n^\varepsilon)$.
Scaling back multiplies the volume by $42^3$.
\end{proof}

\subsection{Packing general convex polyhedra} \label{sec:proof-main}
Since we do not know the dimensions of an optimal container, we try several candidates. For $r\in\{x,y,z\}$, let $W_r=\max_iw_r(K_i)$ be the largest width of an input object in coordinate $r$, and let $L_r^*$ be the corresponding side length of an optimal container. Then
\[
W_r\leq L_r^*\leq nW_r\qquad\text{for }r\in\{x,y,z\}.
\]
The first inequality holds because every object fits into the container. 

For the second, stacking all bounding boxes along one axis gives $L_x^*L_y^*L_z^*\leq nW_xW_yW_z$, which yields $L_r^*\leq nW_r$ using the lower bounds in the other two coordinates. We therefore try the $O(\log^3 n)$ candidate containers with side lengths $L_r\in\{2^jW_r:0\leq j\leq\lceil\log_2n\rceil\}$. At least one candidate satisfies $L_r^*\leq L_r<2L_r^*$ in every coordinate. For each candidate, we scale coordinate $r$ by $1/L_r$ to make the container the unit cube. The scaled objects then have coordinate widths at most $1$, as assumed in \cref{sec:enclose}, and for a candidate satisfying these bounds they can be packed together in the unit cube.

We then compute the enclosures $F_i$, round those with $b_i,c_i\geq1/n$ to $bc$-parallelepipeds, and pack each of their $O(\log^2 n)$ types with \cref{alg:canonical-packing}. For such a candidate, each type is packed into a box of volume $O_\varepsilon(n^\varepsilon)$ by \cref{lem:type-cost}. We handle the remaining objects separately, combine all boxes, and undo the scaling, returning the smallest-volume packing over all candidates. The four steps for a candidate are as follows.

\begin{enumerate}[1.]
\item \emph{Rounded objects.} For each $K_i$ whose enclosure has $b_i,c_i\geq1/n$, we compute the $bc$-parallelepiped $P_i$ and pack each type separately with \cref{alg:canonical-packing}.
\item \emph{Flat objects.} We pack each $K_i$ whose enclosure has $c_i<1/n$ into its own bounding box.
\item \emph{Thin objects.} We fan-pack the remaining enclosures, which have $b_i<1/n\leq c_i$, as in \cref{alg:canonical-packing}, with one object per block.
\item \emph{Combining.} We combine the boxes from these steps into a single box using \cref{lem:combine}.
\end{enumerate}
\Cref{alg:main} summarizes the algorithm.

\begin{algorithm}[h]
\caption{Packing convex polyhedra}
\label{alg:main}
\begin{algorithmic}[1]
\State Let $W_r=\max_iw_r(K_i)$ for $r\in\{x,y,z\}$.
\For{each candidate container with side lengths $L_r\in\{2^jW_r:0\leq j\leq\lceil\log_2n\rceil\}$}
    \State Scale coordinate $r$ by $1/L_r$ and compute $F_i$.
    \State Compute $P_i$ for $b_i,c_i\geq1/n$; pack each type with \cref{alg:canonical-packing}. \Comment{Step 1}
    \State Pack $K_i$ with $c_i<1/n$ in their bounding boxes. \Comment{Step 2}
    \State Fan-pack $F_i$ with $b_i<1/n\leq c_i$, one per block. \Comment{Step 3}
    \State Combine the boxes using \cref{lem:combine}. \Comment{Step 4}
    \State Undo the scaling.
\EndFor
\State \Return the packing of smallest volume.
\end{algorithmic}
\end{algorithm}
\FloatBarrier

The combining step uses the following lemma.

\begin{lemma}
\label{lem:combine}
Let $0<s\leq1$ and $h>0$. Suppose that objects whose $x$- and $y$-widths are at most $1-s$ and whose $z$-widths are at most $h$ are packed by translations into boxes of total volume $V$. In polynomial time, they can be repacked by translations into the box $[0,1]\times[0,1]\times[0,64V/s^2+16h]$.
\end{lemma}

\input{figures/bin-strip}

\begin{proof}
We subdivide each box into cells of length $s$ in the $x$- and $y$-directions and $h$ in the $z$-direction, and assign each object to the cell containing the corner of its bounding box with the smallest coordinates. We extend each cell by $1-s$ in the $x$- and $y$-directions and by $h$ in the $z$-direction, and take its intersection with the original box. Each resulting box contains all objects assigned to its cell and has a base fitting into $[0,1]^2$ and height at most $2h$; see \cref{fig:bin-strip}.

Fix a point $p=(x,y,z)$ in an original box. An extended cell containing $p$ must have its lower corner in $[x-1,x]\times[y-1,y]\times[z-2h,z]$, since its $x$- and $y$-widths are at most $1$ and its $z$-width at most $2h$. The lower corners lie on a grid with spacings $s,s,h$, giving at most $1/s+1\leq2/s$ choices for each of the first two coordinates and at most $3\leq4$ for the third. Thus $p$ lies in at most $16/s^2$ extended cells. Summing the cell volumes counts each point once for every cell containing it, so applying this bound in each original box gives total volume at most $16V/s^2$.

To pack the extended cells, we use the classical strip packing result of Li and Cheng~\cite{LiC90}, as stated by Kar, Khan, and Rau~\cite{KarKR25}: axis-parallel boxes with bases fitting into $[0,1]^2$, heights at most $h'$, and total volume $W$ can be packed by translations in polynomial time into $[0,1]^2\times[0,4W+8h']$. Packing each extended cell together with its assigned objects therefore gives height at most
\[
4\cdot\frac{16V}{s^2}+8\cdot2h=\frac{64V}{s^2}+16h.
\]
\end{proof}

\subsection{Proof of \texorpdfstring{\cref{thm:main}}{Theorem 1}}

\begin{proof}[Proof of \cref{thm:main}]
Every packed object contains the corresponding input polyhedron and all placements are translations, so \cref{alg:main} returns a translational packing. To bound its volume, it suffices to consider a candidate satisfying $L_r^*\leq L_r<2L_r^*$ in every coordinate. After scaling, the optimal packing fits inside the unit cube, and undoing the scaling multiplies volumes by $L_xL_yL_z<8\operatorname{OPT}_{\mathrm{vol}}$. We now bound the volume of the boxes produced in Steps 1--3 for the scaled objects.

In Step 1, \cref{lem:type-cost} shows that each of the $O(\log^2 n)$ types is packed into a box of volume $O_\varepsilon(n^\varepsilon)$. In Step 2, each object fits into a box of dimensions $1\times1\times\frac1n$, and these at most $n$ boxes can be stacked into the unit cube.

In Step 3, the $yz$-shadow of $F_i$ is a parallelogram with edges $(b_i,0)$ and $(q_i,c_i)$, where $b_i<1/n$. As in \cref{alg:canonical-packing}, we order the enclosures by increasing slope $q_i/c_i$ and translate them so that the bottom edges of their shadows lie in the plane $z=0$ and start at $y$-coordinates $0,1/n,2/n,\ldots$. Since the slopes increase with these $y$-coordinates, the shadows have disjoint interiors, so the enclosures do as well. All edge coordinates of the $F_i$ are at most $7$ in absolute value, so the enclosures fit into a box with $x$-extent at most $21$, $y$-extent at most $n\cdot\frac1n+14=15$, and $z$-extent at most $7$, which has volume $O(1)$.

The boxes of Steps 1--3 therefore have total volume $O_\varepsilon(n^\varepsilon\log^2 n)=O_\varepsilon(n^{2\varepsilon})$. Since all packed objects have coordinate widths $O(1)$, we scale by a constant to make these widths at most $1/2$ and apply \cref{lem:combine} with $s=h=1/2$ in Step 4. This gives a box with base $[0,1]^2$ and height $O_\varepsilon(n^{2\varepsilon})$, so undoing this constant scaling gives one box of volume $O_\varepsilon(n^{2\varepsilon})$. Undoing the original coordinate scaling then multiplies this volume by less than $8\operatorname{OPT}_{\mathrm{vol}}$, so the returned packing has volume $O_\varepsilon(n^{2\varepsilon})\cdot\operatorname{OPT}_{\mathrm{vol}}$, and replacing $\varepsilon$ by $\varepsilon/2$ gives the claimed bound. Finally, there are $O(\log^3 n)$ candidates, each step for a candidate takes polynomial time, and no step depends on $\varepsilon$, so the same polynomial-time algorithm achieves the bound for every $\varepsilon>0$.
\end{proof}

We conclude with a remark on the running time. The enclosures $F_i$ are computed only once, since scaling the coordinates maps them to enclosures of the scaled objects. For each of the $O(\log^3(n))$ candidates, all remaining steps take $O(n\log n)$ time, apart from the algorithm of Li and Cheng in \cref{lem:combine}, which is applied to at most $n$ boxes.

\section{Bin packing and strip packing}
\label{sec:bin-strip}

We now prove \cref{thm:bin-strip}. In translational bin packing, the objects must be packed by translations into as few unit cubes as possible. In translational strip packing, they must be packed into the strip $[0,1]^2\times[0,\infty)$, minimizing the height of the packing.

\begin{proof}[Proof of \cref{thm:bin-strip}]
Let $\operatorname{OPT}_{\mathrm{bin}}$ be the minimum number of unit cubes and $\operatorname{OPT}_{\mathrm{strip}}$ the minimum height. Placing the unit cubes of an optimal bin packing next to each other gives a packing into a box of volume $\operatorname{OPT}_{\mathrm{bin}}$, and an optimal strip packing is itself a packing into a box of volume $\operatorname{OPT}_{\mathrm{strip}}$. Hence $\operatorname{OPT}_{\mathrm{vol}}\leq\operatorname{OPT}_{\mathrm{bin}}$ and $\operatorname{OPT}_{\mathrm{vol}}\leq\operatorname{OPT}_{\mathrm{strip}}$, and the $\alpha$-approximation algorithm gives a packing into a box of volume $V\leq\alpha\operatorname{OPT}_{\mathrm{vol}}$.

We first consider strip packing. Let $w_r$ be the largest $r$-width of an object, so that $w_x,w_y\leq1-\delta$. \Cref{lem:combine}, applied to this box with $s=\delta$ and $h=w_z$, gives a packing into the strip of height at most $64V/\delta^2+16w_z$. Every object fits into an optimal strip packing, so $w_z\leq\operatorname{OPT}_{\mathrm{strip}}$, and the height is at most $64\alpha\operatorname{OPT}_{\mathrm{strip}}/\delta^2+16\operatorname{OPT}_{\mathrm{strip}}\leq80\delta^{-2}\alpha\cdot\operatorname{OPT}_{\mathrm{strip}}$.

For bin packing, we cut this strip packing, of height $H$, into slabs of length $\delta$ in the $z$-coordinate and assign each object to the slab containing its lowest point. Since now also $w_z\leq1-\delta$, each object lies in its slab extended by $1-\delta$, which is a box with base $[0,1]^2$ and height $1$. We put each extended slab into its own unit cube. Stacking the unit cubes of an optimal bin packing gives a strip packing, so $\operatorname{OPT}_{\mathrm{strip}}\leq\operatorname{OPT}_{\mathrm{bin}}$. Since $\operatorname{OPT}_{\mathrm{bin}}\geq1$, this uses at most $H/\delta+1\leq81\delta^{-3}\alpha\cdot\operatorname{OPT}_{\mathrm{bin}}$ unit cubes.
\end{proof}

\section{Concluding remarks}
\label{sec:conclusion}

Although our packing algorithm is simple, its analysis relies on recent Kakeya estimates, and it remains open whether the same approximation guarantee can be proved using more elementary arguments. Using higher-dimensional Kakeya estimates for packing is a natural research direction, but arranging the higher-dimensional shadows poses a major obstacle, as it would require new ideas beyond the fan packing we use in two dimensions.

\section*{AI Methodology}
We initially designed the algorithm with a completely different analysis in mind, which did not work out. We then asked ChatGPT~5.5 to repair this analysis. It did not succeed, but its reasoning traces showed that it was looking at the paper of Wang and Zahl~\cite{WangZahl2025}, which made us aware of the connection to the Kakeya problem. Building on this, we developed the proof presented here. We also used GPT Astra 6 and Claude Opus 5.5 for literature search, for creating the TikZ figures, for assistance with language formulations, and for working out details based on the authors' insights and general proof strategies. They also found the exact connection between the Kakeya maximal estimate in~\cite{OpenAILocalSmoothing2026} and the tube-union bound needed for our analysis.

\bibliographystyle{plainurl}
\bibliography{references}
\appendix

\section{The tube-union estimate}
\label{app:tubes}

\begin{theorem}[Wang--Zahl {\cite[Theorem~1.2]{WangZahl2025}}]
\label{thm:wang-zahl}
For all $\varepsilon>0$, there exists $K>1$ so that the following holds for all $\delta>0$ sufficiently small. Let $\mathcal T$ be a set of $\delta$-tubes contained in the unit ball in $\mathbb R^3$, and suppose that every rectangular prism of dimensions $a\times b\times 2$ contains at most $100ab\delta^{-2}$ tubes from $\mathcal T$ (this is true, for example, if the tubes in $\mathcal T$ point in $\delta$-separated directions). For each $T\in\mathcal T$, let $Y(T)\subset T$ be a measurable set with $|Y(T)|\geq\lambda|T|$. Then
\[
\Big|\bigcup_{T\in\mathcal T}Y(T)\Big|\geq\delta^\varepsilon\lambda^K\sum_{T\in\mathcal T}|T|.
\]
\end{theorem}

Taking $Y(T)=T$ and $\lambda=1$ gives a bound on the union of the tubes in which the exponent $K$ plays no role. We then remove the restrictions on the tube locations and on $\delta$.

\begin{proof}[Proof of \cref{thm:tube-union}]
Fix $\varepsilon>0$ and choose $0<\delta_\varepsilon\leq\frac1{10}$ so that \cref{thm:wang-zahl} holds for all $0<\delta<\delta_\varepsilon$. The tubes in $\mathcal T=\{T_\delta(v_i,x_i):1\leq i\leq m\}$ point in $\delta$-separated directions, since the unoriented angle between $v_i$ and $v_j$ is at least $\delta$ for $i\neq j$.

For $\delta<\delta_\varepsilon$, we apply \cref{thm:wang-zahl} locally to remove the unit-ball restriction. We partition the space into cubes of side $1/4$ and let $\mathcal T_Q$ consist of the tubes whose segment midpoints lie in a cube $Q$. Every point of such a tube is within distance $\frac{\sqrt3}{8}+\frac12+\delta<1$ of the centre of $Q$, so all tubes in $\mathcal T_Q$ lie in the unit ball centred there. The theorem therefore applies to each $\mathcal T_Q$ with $Y(T)=T$ and $\lambda=1$.

Each point $x$ lies in at most $9^3$ of these unit balls, since their centres must lie in $x+[-1,1]^3$, which contains at most $9^3$ points of the grid of spacing $1/4$. Consequently,

\[
\begin{aligned}
\operatorname{vol}\!\left(\bigcup_{T\in\mathcal T}T\right)
&\geq\frac1{9^3}\sum_Q
  \operatorname{vol}\!\left(\bigcup_{T\in\mathcal T_Q}T\right)\\
&\geq\frac{\delta^\varepsilon}{9^3}
  \sum_Q\sum_{T\in\mathcal T_Q}\operatorname{vol}(T)
=\frac{\delta^\varepsilon}{9^3}
  \sum_{T\in\mathcal T}\operatorname{vol}(T).
\end{aligned}
\]

For $\delta\geq\delta_\varepsilon$, we bound the number of tubes by placing a cap of angular radius $\delta_\varepsilon/2$ around each $v_i$ on $S^2$. The caps have disjoint interiors, since the angles between the $v_i$ are at least $\delta_\varepsilon$, and each has area $4\pi\sin^2(\delta_\varepsilon/4)\geq\delta_\varepsilon^2/\pi$, using $\sin x\geq2x/\pi$ for $0\leq x\leq\pi/2$. Comparing their total area with the sphere's area $4\pi$ gives at most $N=4\pi^2\delta_\varepsilon^{-2}$ tubes. Since their union contains the largest tube and $\delta^\varepsilon\leq1$, we obtain
\[
\operatorname{vol}\!\left(\bigcup_{T\in\mathcal T}T\right)
\geq\frac1N\sum_{T\in\mathcal T}\operatorname{vol}(T)
\geq\frac{\delta^\varepsilon}{N}\sum_{T\in\mathcal T}\operatorname{vol}(T).
\]
Taking $C_\varepsilon=\max\{9^3,N\}=O(\delta_\varepsilon^{-2})$ proves \cref{thm:tube-union}. Thus $C_\varepsilon$ depends on $\varepsilon$ only through $\delta_\varepsilon$.
\end{proof}

\section{Polylogarithmic bounds}
\label{sec:polylog}

The factor $n^\varepsilon$ in \cref{thm:main} arises solely from the $\delta^\varepsilon$ loss in \cref{thm:tube-union}. To prove \cref{cor:polylog}, we replace this estimate with the tube bounds of OpenAI~\cite{OpenAILocalSmoothing2026,OpenAILean2026}. The preprint claims a loss of $\log^4(1/\delta)$~\cite[Corollary~9.1]{OpenAILocalSmoothing2026}, whereas the formalization verifies a bound with a loss of $\log^{66}(1/\delta)$. We show that these bounds yield approximation ratios $O(\log^7n)$ and $O(\log^{69}n)$, respectively.

\subsection{The tube estimates}

\begin{theorem}[{\cite[\texttt{tube\_union\_logarithmic}]{OpenAILean2026}}]
\label{thm:lean-tubes}
There is a constant $c_0>0$ such that, for all $0<\delta\leq1/e$, all $v_1,\ldots,v_m\in S^2$ with $\|v_i-v_j\|\geq\delta$ for $i\neq j$, and all $x_1,\ldots,x_m\in\mathbb R^3$,
\[
\operatorname{vol}\!\left(\bigcup_{i=1}^mT_\delta(v_i,x_i)\right)\geq\frac{c_0\,m\delta^2}{(1+\log(1/\delta))^{66}}.
\]
\end{theorem}

The formalization states this bound in $\mathbb R^n$, with $m\delta^{n-1}$ in the numerator and exponent $2n(2n+5)$, which equals $66$ for $n=3$; this statement does not seem to appear in~\cite{OpenAILocalSmoothing2026}. As we will show, it yields \cref{thm:tube-union} with the factor $\delta^\varepsilon/C_\varepsilon$ replaced by $C\log^{-66}(1/\delta)$ for some absolute constant $C>0$. Here we absorb the additive $1$ in the logarithmic factor into $C$, using $\delta\leq1/e$.

\Cref{thm:lean-tubes} requires the Euclidean distance between any two direction vectors to be at least the tube radius, whereas our angular separation guarantees only $\|v_i-v_j\|\geq2\sin(\delta/2)\geq2\delta/\pi$. We therefore replace each $T_\delta(v_i,x_i)$ by $T_{2\delta/\pi}(v_i,x_i)$, reducing its radius to $2\delta/\pi$. The direction vectors are separated by at least this new radius, so these replacement tubes satisfy the hypothesis of \cref{thm:lean-tubes}. Each replacement tube lies inside its original tube, so a lower bound on the volume of their union also bounds the volume of the original union from below.

Since changing the radius by a constant factor only affects the constants, \cref{thm:lean-tubes} gives the desired bound with factor $C\log^{-66}(1/\delta)$.

The preprint~\cite{OpenAILocalSmoothing2026} instead claims a Kakeya maximal estimate, formulated using the Kakeya maximal function $K_\delta f\colon S^2\to[0,\infty]$, defined for a locally integrable function $f\colon\mathbb R^3\to\mathbb R$ by
\[
K_\delta f(\omega)=\sup_{x\in\mathbb R^3}\frac{1}{\operatorname{vol}(T_\delta(\omega,x))}\int_{T_\delta(\omega,x)}|f|.
\]
Thus $K_\delta f(\omega)$ is the supremum of the averages of $|f|$ over tubes of length $1$ and radius $\delta$ in direction $\omega$. For example, if $E$ is a union of such tubes and $f=\mathbf 1_E$, then $K_\delta f(\omega)=1$ whenever one of the tubes has direction $\omega$. For nearby directions, a tube still lies largely in $E$, so $K_\delta f$ remains close to $1$. We measure $K_\delta f$ and $f$ using the norms
\[
\|K_\delta f\|_{L^3(S^2)}=\left(\int_{S^2}(K_\delta f(\omega))^3\,d\omega\right)^{1/3}
\quad\text{and}\quad
\|f\|_{L^3(\mathbb R^3)}=\left(\int_{\mathbb R^3}|f(x)|^3\,dx\right)^{1/3}.
\]
We write $L^3(\mathbb R^3)$ for the space of functions with finite $L^3$ norm. For an indicator function $f=\mathbf 1_E$, this norm is $\|f\|_{L^3(\mathbb R^3)}=\operatorname{vol}(E)^{1/3}$.

\begin{theorem}[{\cite[Corollary~9.1]{OpenAILocalSmoothing2026}}]
\label{thm:kakeya-maximal}
There is a constant $C_K\geq1$ such that, for all $0<\delta\leq1$ and all $f\in L^3(\mathbb R^3)$,
\[
\|K_\delta f\|_{L^3(S^2)}\leq C_K\log^{4/3}(2/\delta)\,\|f\|_{L^3(\mathbb R^3)}.
\]
\end{theorem}

The preprint states this estimate in $\mathbb R^n$ for every $n\geq2$, with the $L^n$ norms and the power $\frac4{p_c}$, where $p_c=\frac{2n}{n-1}$; for $n=3$, this is the power $\frac43$ above.

\begin{lemma}
\label{lem:maximal-to-tubes}
If \cref{thm:kakeya-maximal} holds, then, for $0<\delta\leq1/e$, the logarithmic factor in \cref{thm:lean-tubes} can be replaced by $\log^{-4}(1/\delta)$.
\end{lemma}

\begin{proof}
Let $0<\delta\leq1/e$ and let $v_1,\ldots,v_m$ and $x_1,\ldots,x_m$ be as in \cref{thm:lean-tubes}. Write $E=\bigcup_{i=1}^mT_\delta(v_i,x_i)$ and $f=\mathbf 1_E$, and for each $i$ define the cap of angular radius $\delta/3$ centred at $v_i$ by
\[
B_i=\Bigl\{\omega\in S^2:\arccos(\omega\cdot v_i)\leq\frac\delta3\Bigr\}.
\]

These caps are disjoint. Indeed, suppose that $\omega$ lies in both $B_i$ and $B_j$. Angles between unit vectors satisfy the triangle inequality, so the angle between $v_i$ and $v_j$ is
\[
\arccos(v_i\cdot v_j)\leq\arccos(\omega\cdot v_i)+\arccos(\omega\cdot v_j)\leq\frac{2\delta}3.
\]
Since a chord of the unit circle is shorter than the corresponding arc, $\|v_i-v_j\|\leq\arccos(v_i\cdot v_j)\leq\frac{2\delta}3<\delta$, contradicting the assumption $\|v_i-v_j\|\geq\delta$.

A cap of angular radius $\alpha$ has area $2\pi(1-\cos\alpha)=4\pi\sin^2\frac\alpha2$. Since $\sin x\geq\frac{2x}\pi$ for $0\leq x\leq\frac\pi2$, each $B_i$ has area at least
\[
4\pi\sin^2\frac\delta6\geq4\pi\Bigl(\frac{\delta}{3\pi}\Bigr)^2=\frac{4\delta^2}{9\pi}\geq\frac{\delta^2}8.
\]

We claim that $K_\delta f(\omega)\geq\frac18$ for every $\omega\in B_i$. Let $S=x_i+\frac12T_\delta(\omega,0)$. Then $S\subseteq T_\delta(\omega,x_i)$ and $\operatorname{vol}(S)=\frac18\operatorname{vol}(T_\delta(\omega,x_i))$. Moreover, $S\subseteq T_\delta(v_i,x_i)\subseteq E$. Indeed, let $x_i+t\omega+e$ be a point of $S$, with $|t|\leq\frac14$ and $\|e\|<\frac\delta2$. Since $\|\omega-v_i\|\leq\arccos(\omega\cdot v_i)\leq\frac\delta3$, its distance to the point $x_i+tv_i$ of the axis segment of $T_\delta(v_i,x_i)$ is at most
\[
|t|\,\|\omega-v_i\|+\|e\|<\frac14\cdot\frac\delta3+\frac\delta2<\delta.
\]
Hence $E$ covers at least $\frac18$ of the tube $T_\delta(\omega,x_i)$, which gives $K_\delta f(\omega)\geq\frac18$.

We now bound $\|K_\delta f\|_{L^3(S^2)}^3$ from both sides:
\begin{align*}
\frac{m\delta^2}{8^4}
&\leq\sum_{i=1}^m\int_{B_i}(K_\delta f(\omega))^3\,d\omega
&&\text{since $K_\delta f\geq\tfrac18$ on $B_i$ and $B_i$ has area at least $\tfrac{\delta^2}8$,}\\
&\leq\int_{S^2}(K_\delta f(\omega))^3\,d\omega
&&\text{since the caps $B_i$ are disjoint,}\\
&=\|K_\delta f\|_{L^3(S^2)}^3
&&\text{by the definition of the norm,}\\
&\leq C_K^3\log^4(2/\delta)\,\|f\|_{L^3(\mathbb R^3)}^3
&&\text{by \cref{thm:kakeya-maximal},}\\
&=C_K^3\log^4(2/\delta)\operatorname{vol}(E)
&&\text{since $f=\mathbf 1_E$.}
\end{align*}
Rearranging gives the claimed bound with $c_0=\frac{1}{16\cdot8^4C_K^3}$.
\end{proof}

\subsection{Proof of \texorpdfstring{\cref{cor:polylog}}{Corollary 2}}

\begin{proof}[Proof of \cref{cor:polylog}]
We follow the proof of \cref{thm:main}, but use \cref{thm:lean-tubes} in place of \cref{thm:tube-union}. By the remark after \cref{thm:lean-tubes}, this replaces the factor $\delta^\varepsilon/C_\varepsilon$ by $C\log^{-66}(1/\delta)$. We trace this change through the three steps of the analysis.

First, \cref{thm:counting} holds with $C_\varepsilon n^\varepsilon$ replaced by $O(\log^{66}n)$. Indeed, the volume bound of \cref{sec:lower-bound}, with $\delta=\frac{b\tau}{1000}$ and the new factor, shows that at most
\[
O\Bigl(\frac{\log^{66}(1000/(b\tau))}{bc\tau}\Bigr)
\]
objects have $a_i<\frac\tau{100}$. If $b\tau\geq n^{-2}$, then $\log\frac{1000}{b\tau}\leq\log(1000n^2)=O(\log n)$, so this number is $O\bigl(\frac{\log^{66}n}{bc\tau}\bigr)$; if $b\tau<n^{-2}$, then $|\mathcal P|\leq n<\frac1{b\tau}\leq\frac1{bc\tau}$ anyway. Together with the at most $\frac{100}{bc\tau}$ objects with $a_i\geq\frac\tau{100}$, this gives $|\mathcal P|=O\bigl(\frac{\log^{66}n}{bc\tau}\bigr)$.

Second, with this bound, the computation in the proof of \cref{lem:mst} gives
\[
bc\sum_k\operatorname{MST}_k\leq1+O(\log^{66}n)\ln(2n)=O(\log^{67}n),
\]
so, by the proof of \cref{lem:type-cost}, each type is packed into a box of volume $O(\log^{67}n)$.

Finally, there are $O(\log^2n)$ types, so the boxes of Steps 1--3 in the proof of \cref{thm:main} have total volume $O(\log^{69}n)$. The rest of that proof is unchanged, so the returned packing has volume $O(\log^{69}n)\cdot\operatorname{OPT}_{\mathrm{vol}}$.

If \cref{thm:kakeya-maximal} holds, then by \cref{lem:maximal-to-tubes} the same argument works with the exponent $66$ replaced by $4$ throughout, which gives the approximation ratio $O(\log^7n)$.
\end{proof}

\end{document}